\documentclass[%
 reprint,
 amsmath,amssymb,
 aps,
]{revtex4-2}

\usepackage{graphicx}
\usepackage{dcolumn}
\usepackage{bm}
\usepackage{hyperref}

\begin{document}

\preprint{APS/123-QED}

\title{My model, it has three layers: a reduced model of the smectic transition in two dimensions}

\author{David A. King}
\email{daviking@sas.upenn.edu}
\affiliation{Department of Physics and Astronomy, University of Pennsylvania, 209 South 33rd St., Philadelphia, PA, 19104.
}
\date{\today}

	\begin{abstract}
   \begin{center}
    \noindent \textit{My model, it has three layers, \\
Three layers is nematic. \\
And had it just two layers, \\
it would be a smectic}		
  \\
\end{center}

  We study a reduced model of the smectic transition in two dimensions where the particles occupy three equally spaced layers. The role of particle geometry comes in through the interactions between particles on the central layer and those above and below. The system is understood to be smectic when the central layer is empty, and nematic when all three are equally occupied. It is possible to compute the free energies of these states exactly. We find that the free energy of the nematic can only exceed that of the smectic if the particle tips are sufficiently wide, mirroring the fact that ellipsoids do not make a smectic but sphero-cylinders do. 
	\end{abstract}
\maketitle
\onecolumngrid
	\section{Introduction}
	\label{sec:Intro}
Four score and seven years ago, Lewi Tonks brought forth a new model \cite{Tonks1936TheSpheres}. This was a one-dimensional gas of identical particles interacting by excluded volume and it now bears his name; the Tonks gas. He provided the exact solution for the partition function, equation of state and various other properties of his gas. What he learned from this investigation he applied to develop equations of state for two- and three-dimensional gasses. This kind of development is common in physics. The degrees of freedom in a complicated physical system are reduced, simplifying its mathematical analysis and allowing techniques and intuitions to be uncovered which are brought to bear on more realistic systems.

Exactly soluble models of liquid crystals are hard to come by. One of the most famous examples is due Onsager \cite{Onsager1949TheParticles}. He considered how suspensions of long and thin particles transition from an isotropic state, where both the positions of the particles and the directions of their long axes are uniformly distributed, to a nematic state, in which the long axes pick up a preferential direction while the positions remain disordered. The transition boils down to a trade-off between translational and orientational entropy. The isotropic phase has greater orientational entropy but at the cost of reduced translational entropy, since two perpendicular particles mutually exclude more volume than if they were parallel. The nematic, on the other hand, pays in orientational entropy to gain translational entropy. Onsager's understanding not only put forward an important framework which has been used to study a variety of ordering transitions \cite{Mulder1987Density-functionalFluid, Frenkel1991,Drwenski2016,Blaak2004,Mulder1989}, but also a geometric criterion for nematic formation; the particles need to be at least four times longer than wide and all other geometric features are irrelevant. 

The smectic-A liquid crystal adds one-dimensional positional order to the nematic. In this phase, the particles lie in parallel layers with their long axes aligned with the layer normals. Smectics are more challenging than nematics. Not only does no exactly soluble model exist, but the particle geometry appears to play a significantly more subtle role in their formation. For example, parallel spheroids form a smectic phase but ellispoids do not \cite{Frenkel1991,Frenkel1984,Stroobants1986}. Yet more evidence of a small change in particle architecture eliminating the smectic phase is found with $N$-CB molecules \cite{Gray1976,Cacelli2007}. These may be thought of as having a ``body'' and a ``tail''. The body is made from two aromatic rings with a nitrile group on one end. The tail is attached to the other and is an $N$ carbon long alkyl group. Here, increasing $N$ from $5$ to $8$ allows the smectic to emerge. 

The common feature separating $5$-CB from $8$-CB and ellipsoids from sphero-cylinders is the effective width of the particle tips. The tip of an ellipsoid is more pointed than the hemisphere capping a sphero-cylinder and hence it excludes less volume. The same is true for the shorter tail of $5$-CB; the restrictions it places on the tails of neighboring particles are lesser than those imposed by $8$-CB. Recently, the effect this has on smectic formation was described in the context of a 2D mean-field theory \cite{King2023}. The essence of that analysis was that smectics formed by sphero-cylinders and $8$-CB are more stable to fluctuations trying to restore the nematic, because the spaces between their layers are harder to occupy. This reasoning led to a geometric criterion for smectic formation. Simply put, the tip must be sufficiently wide. However, this treatment made various simplifications and approximations, beyond working in only two-dimensions, the impacts of which are not immediately clear. 

In this paper, we study a reduced model of a two dimensional smectic which may be analysed exactly in the thermodynamic limit. This is made possible by formulating it as an embellished Tonks gas. We suppose there are three equally spaced layers the particles may occupy. The outer two represent those of a smectic and that in the centre represents an interstitial layer. A smectic phase is indicated by the central layer emptying, while a nematic phase has each layer equally occupied. The particles only interact by excluded volume and the layer spacing is chosen such that a particle on the central layer hinders those on the outer layers by colliding its tip with theirs. The outer layers are taken to be sufficiently well separated that they do not interact. We show that, depending on the concentration and the particle geometry, the free energy of the nematic can exceed the that of the smectic, implying the smectic becomes thermodynamically favoured. Consistent with the previous mean-field model, we find that the particle tip needs to be sufficiently wide in order for a smectic to form. This places those intuitive arguments on firmer footing and allows the smectic transition in this model to be understood as an entropic trade-off analogous to the isotropic-nematic transition. 

Our model and its analysis are reminiscent of many other quasi-one-dimensional embellishments of Tonks' gas. A notable early example is Barker's tunnel fluid \cite{Barker1961, Barker1962}, intended as an improvement to the cell model of fluids by Lennard-Jones and Devonshire \cite{LennardJones1937, Barker1955} and another example of the ease of analysis afforded by lowering the number of dimensions providing insight for more realistic systems. Whereas the cell model had each particle confined to a cage by its surroundings, Barker saw them in chains, compelled to lie along narrow, almost one-dimensional, tunnels. Percus and Zhang \cite{Percus1990} were the first to suggest that, as the width of the tunnel was increased, some kind of ordering may take place. While we can rule out a true phase transition with the help of van Hove \cite{VanHove1950}, Gurin et. al. \cite{Gurin2018,Gurin2015} indeed show that a mono-layer smoothly gives way to a smectic-like pair. 

Our work can be interpreted as confirming this observation and elucidating the role of particle geometry in the context of a new exact solution. All of the previous quasi-one-dimensional models mentioned rely on the ``transfer operator method'' \cite{Lieb1966, Kofke1993}. Here, the partition function is computed in the thermodynamic limit by determining, usually numerically, the largest eigenvalue of an associated integral equation. In our case, by dividing the system appropriately, we are able to compute the configurational integral directly. 

The mathematical analysis of our model will be quite involved. Therefore, after describing and formulating the model precisely in section \ref{sec:Model}, we shall present an informal assessment of the problem in section \ref{sec:Inform}. We hope this provides an intuitive discussion of the physics underlying the system, before it gets lost in the technical weeds. Section \ref{sec:PF} then gives the calculation of the partition function. This is used to compute the nematic and smectic free energies in section \ref{sec:SmectTran} where we provide a phase diagram showing where, as a function of particle density and geometry, the smectic is favoured over the nematic. We conclude in section \ref{sec:Conc} with a discussion of our results and how the work may be extended. 
 
\section{Model description and problem statement}
\label{sec:Model}
	\begin{figure}\includegraphics[width=16cm]{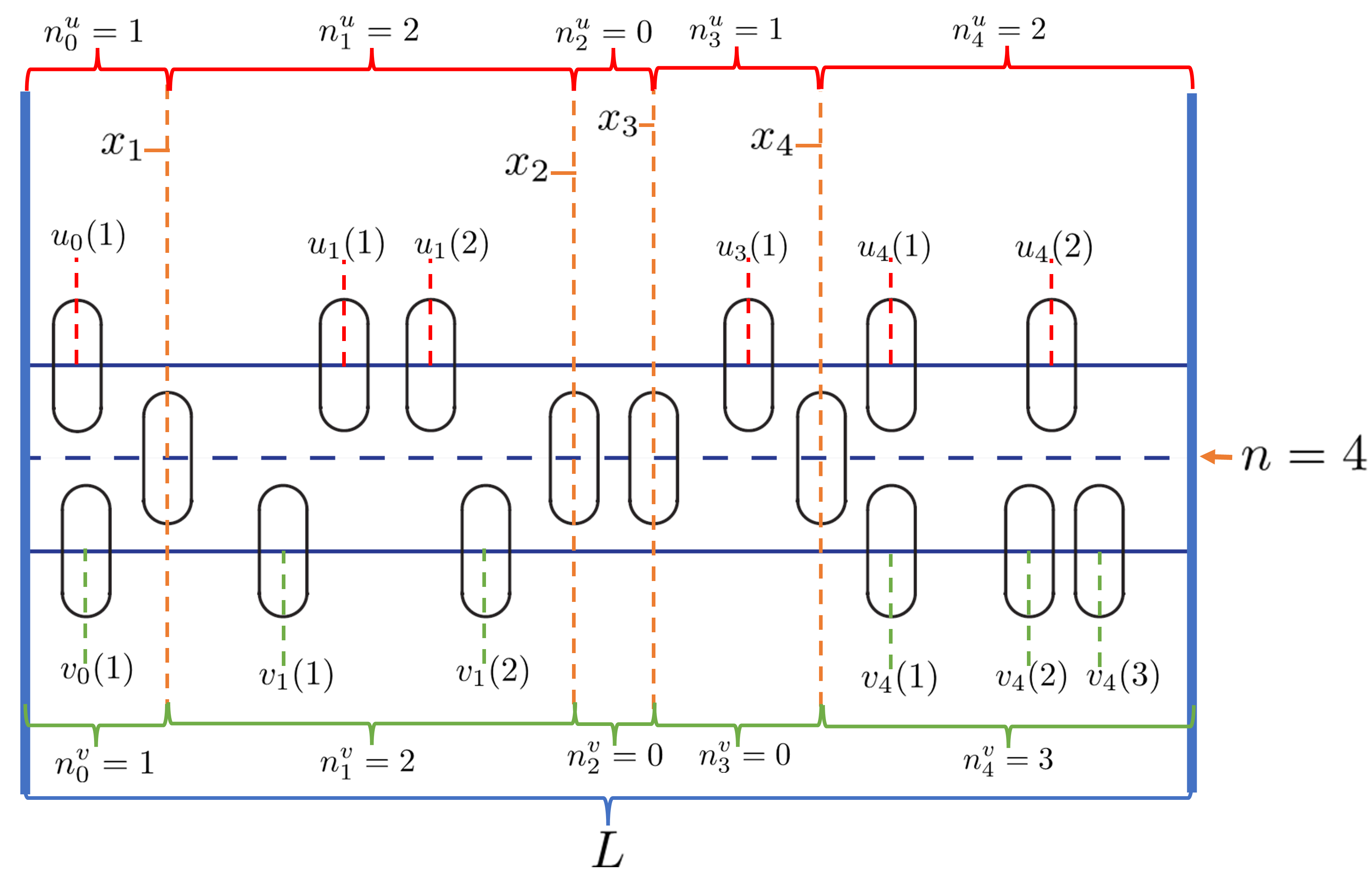}
		\caption{\label{fig:Sys} A sketch of a small system with a total number $N = 16$ of particles labelled according to our prescription. All three layers are the same length, $L$, and have hard walls at their ends. The total number of particles on the central layer is $n$, in this case $n = 4$. These particles are at positions $x_k$ with $k \in [1,n]$, shown in orange. For ease of notation, we suppose the walls are at positions $x_0 = 0$ and $x_{n+1} = 0$. Between the $k^{th}$ and $(k+1)^{st}$ particles on the central layer, there are $n_k^{u}$ particles on the upper layer and $n_k^{v}$ on the lower. For example, between the particle at $x_3$ and $x_4$, we have $n_3^{u} =1$ and $n_3^{v}= 0$. The positions of these particles are labelled $u_k(j)$ with $j \in [1,n_k^{u}]$ for the upper layer and $v_k(j)$ with $j \in [1,n_k^{v}]$ for the lower. The essence of our approach is that the particles on the upper and lower layers form $n+1$ separate Tonks' gasses and they excluded extra length to those particles on the central layer, which may then be treated as their own Tonks gas.}
	\end{figure}
\begin{figure}\includegraphics[width=8cm]{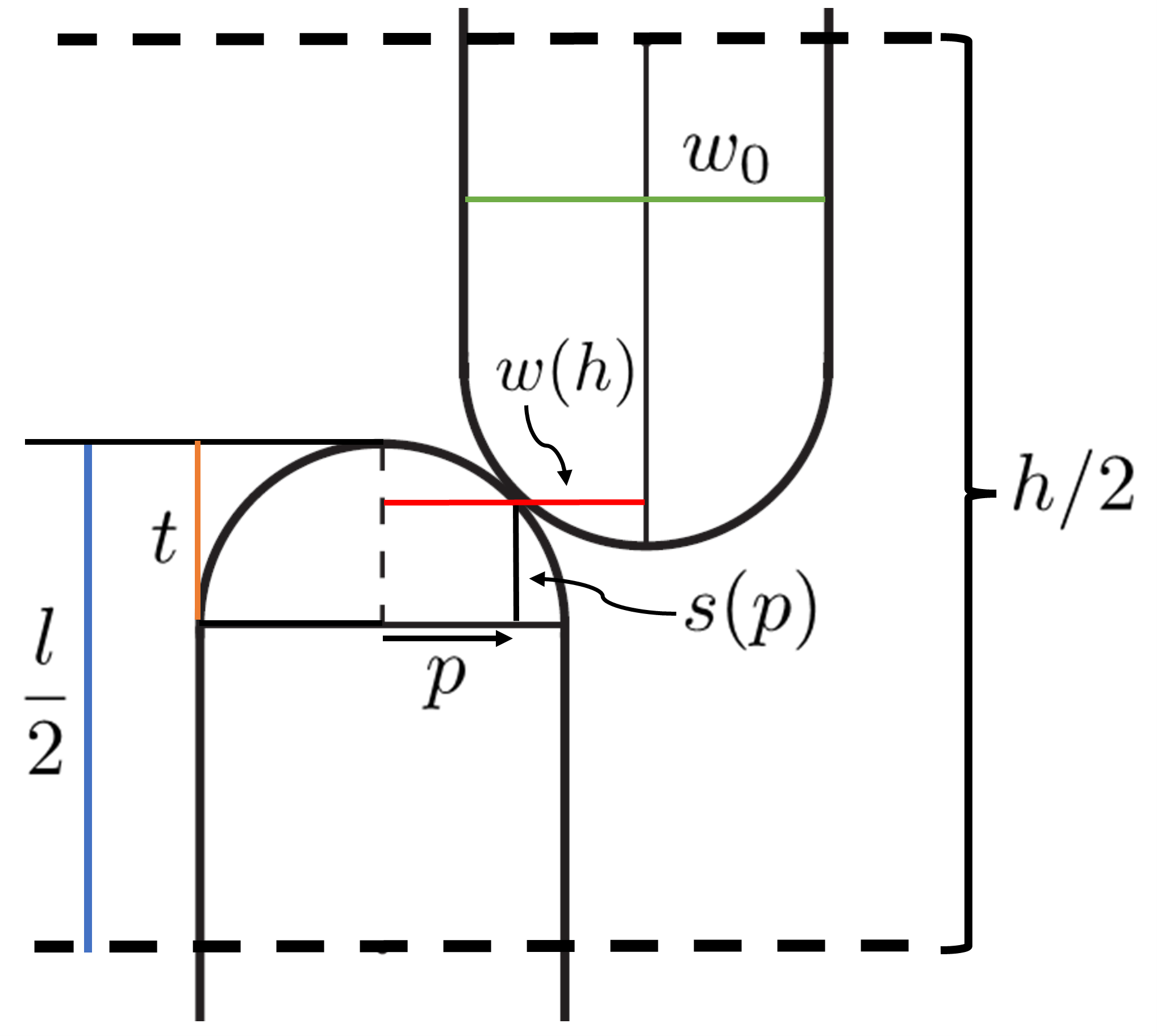}
		\caption{\label{fig:Geom} A sketch showing the geometry of the particles we consider. They have rectangular midsections with width $w_0$, as shown in green. Their tips have a length $t$ (orange) and a total length $l$ (blue). Only half ($l/2$) of the particles are shown, but they are symmetric about their centres. The tip shape is described by the function $s(p)$, measuring the distance from the solid black line to the end of the particle at the point $|p|< w_0/2$ away from the particle centre. The function is assumed symmetric; $s(p) = s(-p)$. The particles are shown as they just come into contact. This happens when their centres are separated by the distance $w(h)$. This distance depends on the geometry of their tips.}
\end{figure}

We consider the system as sketched in Fig.\ref{fig:Sys}. There are $N$ identical particles which interact exclusively by excluded volume. Each may occupy any $x$-position provided they are not overlapping any other particle or the two walls at $x=0$ and $x=L$. The $y$-positions of the the particles are restricted to three discrete layers each separated by $h/2$. The particles are all oriented in the same direction with their long axes perpendicular to the layers. The particles are up-down and left-right symmetric, possess rectangular midsections of width $w_0$ and their length is $l$. We shall consider the case when $l \leq h \leq 2l$. This means that a particle on the central layer may collide with those on the outer layers, but two particles on the outer layers do not interact. Furthermore, we assume that when particles on the central and outer layers collide, they do so at their tips. When this collision happens, the centres of the two colliding particles are separated by $w(h)$ as shown in Fig.\ref{fig:Geom}. This distance of closest approach must depend on the distance between the layers as well as the tip geometry. While we will not consider any concrete tip shapes, $w(h)$ does allow us to compare ``fat'' and ``thin'' tips. A fatter tip has a \textit{larger} $w(h)$ than a thinner tip. For a given particle geometry $w(h)$ may be extracted directly from the shape of the tip. This is described by the function $s(p)$ which gives the height of the tip about the midsection at each point, $|p| < w_0/2$, along the width of the particle. The symmetry of the tip enforces $s(p) = s(-p)$ and, if the height of the tip is $t$, we require $s(0) = t$. It may be shown that 
\begin{equation}
    2 s\left(w(h)/2\right) = h/2 - l + 2t,
\end{equation}
establishing the direct relationship between $s(p)$ and $w(h)$ \cite{King2023}. 
We are interested in the relative occupancy of the three layers. We will use this mark whether the system is nematic or smectic. If all layers are equally occupied the system is nematic. When the central layer is depleted and the outer two are equally occupied, the system is smectic. This assignment is natural if $h$ is identified as the smectic layer spacing because when each layer is equally occupied, the associated smectic order parameter vanishes identically by symmetry. 
	
Our goal is to compare the free energies of the nematic and smectic phases as defined above. This requires computing the partition function of the three layer system as a function of the occupation numbers of the layers. As each layer is one dimensional and is filled with particles interacting by excluded volume, it is natural to expect this calculation to follow that for the Tonks gas closely \cite{Tonks1936TheSpheres}. In that case, the calculation is facilitated by the unique ordering of the particles from left to right. This allowed the configurational integral to be computed by integrating over the positions of each of the particles in turn. In our case, a similar kind of ordering is crucial to make progress.  
	
In the three layer system, we cannot order \textit{all} of the particles from left to right. This is because a particle on the top layer which is to the left of a particle on the bottom may also occupy a position to its right if permitted to do so by the particles in the centre. This situation can be seen in Fig.\ref{fig:Sys}. The particles may, however, be ordered on each layer from left to right. This is the basis for our approach. 
	
Let us call the number of particles on the central layer $n$. The order of these particles from left to right cannot change, so we are permitted to label them with the index $k \in [1, n]$ and write their positions as $x_k$. This means $x_1$ is the position of the left-most particle on the central layer, and $x_n$ the right-most. We can order the particles on the upper and lower layers similarly, but it is more convenient to pursue a different strategy. Between the $k^{th}$ and $(k+1)^{st}$ particles on the central layer there may be $n^{u}_{k} \geq 0$ particles on the upper layer and $n^{v}_{k} \geq 0$ on the lower layer. The order of these two sets of particles cannot be changed and so we can index them similarly. The positions of those particles on the upper layer between $x_k$ and $x_{k+1}$ are $u_k(j)$, with $j \in [1 , n^u_k]$. The equivalent positions for the lower layer are $v_k(j)$, with $j \in [1 , n^v_k]$. By setting $x_0 = 0$ and $x_{n+1} =L$ those particles on the outer layers between the walls and the extremal particles on the central layer may be included by extending the same notation to $k=0$ and $n+1$. In summary, the positions of all of the particles and the walls are written as follows
	\begin{equation}
		\begin{split}
			\text{Upper Layer:} \ \  &u_k(j) \ \ \text{with} \ j \in [1,n^{u}_k] \ \ \text{and} \ \ k \in [0,n],\\
			\text{Central Layer:} \ \ &x_k \ \ \text{with} \ k \in [0,n+1] \\
			\text{Lower Layer:} \ \ &v_k(j) \ \ \text{with} \ j \in [1,n^{v}_k] \ \ \text{and} \ \ k \in [0,n].
		\end{split}
	\end{equation}
	This prescription of the system is sketched in Fig.(\ref{fig:Sys}). Note that, since the total number of particles in the system is $N$, we must have
	\begin{equation}
		\label{eq:Nconst}
		n + \sum_{k=0}^{n} n_k^{u} + n_k^{v} = N. 
	\end{equation}
Let us compile the occupation numbers into a vector,
\begin{equation}
	\textbf{N} = (n,n^u_0 , n^v_0, \cdots, n^u_n,n^v_n),
\end{equation} 
and the positions of particles on each layer into the three vectors
\begin{subequations}
	\begin{equation}
		\textbf{X} = (x_1,\cdots, x_n),
	\end{equation}
	\begin{equation}
		\textbf{U} = \left(u_0(1), \cdots, u_0(n^u_0), u_1(1), \cdots, u_n(n^u_n)\right),
	\end{equation}
	and
	\begin{equation}
		\textbf{V} = \left(v^0_1, \cdots, v_0(n^v_0), v_1(1), \cdots, v_n(n^v_n)\right).
	\end{equation}
\end{subequations}
Using this notation, we can define the partition function for a given $\textbf{N}$ as
\begin{equation}
	\label{eq:PFformal}
	\mathcal{Z}(\textbf{N}) = \int d\textbf{X} d\textbf{U} d\textbf{V} \ \Theta\left(\textbf{X},\textbf{U},\textbf{V}|\textbf{N} \right),
\end{equation}
where $\Theta\left(\textbf{X},\textbf{U},\textbf{V}|\textbf{N} \right)$ is a unit selector function which picks out the allowed particle positions with none overlapping.

We will compute this partition function in the next section, for now let us discuss how it may be used to find the free energy of the nematic and smectic phases. Whether or not the system is in a particular phase is determined by the vector $\textbf{N}$. For example, for the system to be smectic, the number of particles on the central layer, $n$, must be zero. This means all $n_k^{u,v} = 0$ for $k >0$. The non-zero $n_0^{u}$ and $n_0^{v}$ are also required to share the $N$ particles equally. If we define the vector 
\begin{equation}
	\label{eq:NSm}
	\textbf{N}_{\text{Sm}} = (0,N/2,N/2),
\end{equation} 
then a system with $\textbf{N} = \textbf{N}_{\text{Sm}}$ is a smectic and the partition function for the smectic is formally computed from 
\begin{equation}
	\label{eq:ZSm}
	\mathcal{Z}_{\text{Sm}} =  \mathcal{Z}\left(\textbf{N}_{\text{Sm}}\right).
\end{equation} 
From which we get the free energy
\begin{equation}
	\beta F_{\text{Sm}} = - \log\mathcal{Z}_{\text{Sm}},
\end{equation}
where $\beta = (k_B T)^{-1}$. Although the form of this may be anticipated by virtue of the smectic being two non-interacting Tonks gasses, we shall leave the computation of this until later. 

The condition on $\textbf{N}$ for the nematic is less straightforward. In this case we only specify that all of the layers are equally occupied,
\begin{equation}
	\label{eq:NemConst}
	n = N/3, \ \ \text{and} \ \ \sum_{k=0}^{N/3} n_k^{u} = \sum_{k=0}^{N/3} n_k^{v} = N/3,
\end{equation}
all the $n_k^{u}$ and $n_k^{v}$ are otherwise undetermined. In order to determine the partition function for the nematic state, we must therefore sum (\ref{eq:PFformal}) over all $\textbf{N}$ satisfying (\ref{eq:NemConst}), with each configuration weighted appropriately. The appropriate weighting is found by multiplying the total number of nematic states, $\mathcal{N}$, by the proportion, $\mathcal{P}(\textbf{N})$, which have a particular configuration $\textbf{N}$. This weighting is required to take into account the following fact; there are many more ways of distributing the particles on the outer layers with all $n_k^{u,v}$ roughly equal than there are with all but one equalling zero. Note that this weight is not needed for the smectic phase. In that case, aside from re-labelling the particles, there is only one way of realising the smectic state with $\textbf{N}_{\text{Sm}}$. This weighting ensures that, even if the particles do not interact with eachother, the nematic state is favoured by virtue of the many more ways of arranging the particles over three layers than two. 

Once this weighting is taken into account, the nematic partition function is given by
\begin{equation}
	\label{eq:ZNemForm}
	\mathcal{Z}_{\text{N}} = \mathcal{N} \sum_{\textbf{N}}^{*}\mathcal{P}(\textbf{N}) \mathcal{Z}(\textbf{N})
\end{equation}
where the star on the summation indicates that it is subject to the conditions (\ref{eq:NemConst}).
The free energy for the nematic is of course
\begin{equation}
	\beta F_{\text{N}} = - \log \mathcal{Z}_{\text{N}}. 
\end{equation}

These expressions are currently purely formal. Not only do we need to calculate $\mathcal{Z}(\textbf{N})$, but a combinatoric argument is required for $\mathcal{N}(\textbf{N})$ before the constrained sums in (\ref{eq:ZNemForm}) may be taken. These calculations are detailed and will be the focus of much of the rest of the paper. Our goal is to determine for which particle concentrations and shapes should we expect a smectic phase to be favoured over the nematic. In other words, our central question is; ``When is $\beta F_{\text{Sm}} < \beta F_{\text{N}}$?''
	\section{An Informal Assessment of the Problem}
\label{sec:Inform}
Our forthcoming analysis based on the approach outlined above will be rather involved. So before diving into the detail, it is worth briefly discussing the physics in a simple, heuristic context. 

The fundamental physics governing the occupancy of the central layer can be understood entropically and we argue that this is also essential to real, many layer smectic forming systems. Suppose initially that the central layer is empty. We can consider each outer layer separately because they are too far apart for their occupying particles to interact. Within each layer though, the particles exclude some length to their neighbours. This reduces their translational freedom, and thus the system's entropy, from the ideal case. When particles are introduced to the central layer, not only is their own translational freedom reduced by the length that they exclude amongst themselves, their presence also reduces the space available to the particle on the outer layers. Therefore, the entropic cost of adding a particle to the central layer is \textit{higher} than adding one to the outer layers. Emptying the central layer, however, is itself punished entropically. This is because there are many more ways of realising the system with the particles spread evenly between the layers. Here we shall outline a crude model which demonstrates how this trade off can lead to the smectic phase being preferred over the nematic. Our proceeding exact analysis will be compared to this intuitive picture. 

Our goal is to construct heuristic free energies for the nematic and smectic states based on the above physics. In both instances, because the interactions are all hard-core, the free energy is entirely entropic,
\begin{equation}
	\beta F = - S = -\log \Omega,
\end{equation}
where we have used Boltzmann's definition of the entropy to write it in terms of the number of states accessible to the system, $\Omega$. This can be split up into the number of ways of arranging the particles between the layers, $\mathcal{A}$, and the accessible length to each particle, $\mathcal{L}$, so that
\begin{equation}
	\beta F = -\log \mathcal{A} - N \log \mathcal{L}.
\end{equation}
The number of ways of arranging the particle between the layers in the smectic state, $\mathcal{A}_{\text{Sm}}$, is the number of ways of distributing the $N$ evenly between the two outer layers, 
\begin{equation}
	\mathcal{A}_{\text{Sm}} = \binom{N}{N/2}.
\end{equation}
To estimate $\mathcal{L}_{\text{Sm}}$, we suppose that every particle can access the full length of the layer, $L$, minus the total length excluded by the other particles on that layer. This should be proportional to the number of particles on each layer, so we shall write $\mathcal{L}_{\text{Sm}} \sim L - N a/2$. The constant $a$ can be understood as the effective width of the particles. In the limit $N \to \infty$, we have
\begin{equation}
	\beta F_{\text{Sm}} = - N \log 2 - N \log (L - N a/2).
\end{equation}
In the nematic state, $\mathcal{A}_{\text{N}}$ is the number of ways of splitting $N$ particles into three equal piles, one for each layer,
\begin{equation}
	\mathcal{A}_{\text{N}} = \frac{N!}{(N/3!)^3}.
\end{equation}
The entropic cost of adding a particle to the central layer needs to be accounted for in our approximation of $\mathcal{L}_{\text{N}}$. To do this, we suppose that the presence of the particles on the central layer \textit{increases} the effective width of the particles to $b > a$. Now, recalling that the number of particles on each layer is $N/3$ we have, for large $N$,
\begin{equation}
	\beta F_{\text{N}} = - N \log 3 - N \log (L - N b/3).
\end{equation}
The condition that the smectic is favoured over the nematic is $\beta F_{\text{Sm}} < \beta F_{\text{N}}$. This can be written 
\begin{equation}
	\label{eq:CrudeInEq}
	\frac{1 - \rho b/3}{1 - \rho a/2} < \frac{2}{3},
\end{equation}
where $\rho = N/L$ is the number density of the particles. Let us emphasise the meaning of the two sides of this inequality. The left hand side measures the reduction of translation entropy incurred by adding particles to the central layer. To avoid this toll, the system attempts to empty the central layer, but that requires a payment measured by the right hand side. The inequality balances the books between these two entropic expenditures. It can be rearranged to find the value of $b$ for which the smectic is favoured
\begin{equation}
	b > \frac{1}{\rho} + a. 
\end{equation}
How can we estimate the effective widths, $a$ and $b$, in terms of the particle geometry? In the smectic, each particle excludes a width $w_0$; the width of its midsection. Therefore, $a= w_0$. In the nematic, the excluded width is slightly more complicated. Suppose that two particles on the upper layer approach each other while a particle on the central layer sits between them. The particle on the central layer can slide freely, but cannot penetrate those on the upper layers. The closest distance the two upper layer particles can possibly get to each other is $2 w(h)$. Hence we may suppose that $b \sim 2 w(h)$. Our heuristic condition for smecticity is 
\begin{equation}
	2w(h) \gtrsim \frac{1}{\rho} + w_0. 
\end{equation}
This is essentially the same result as obtained by a more detailed mean-field-theory argument\cite{King2023}. Furthermore it indicates that, for a sufficiently wide particle tip, we can expect a smectic state to be favoured over the nematic. This discussion has been intuitive but crude, and it is difficult to tell how accurate it really is. In the following section, we shall set about the program outlined in section \ref{sec:Model}. Eventually, we shall obtain an inequality very similar to (\ref{eq:CrudeInEq}) albeit significantly more complicated. The arguments put forth in this section shall allow us to understand it physically in much the same way. 
\section{Partition function}
\label{sec:PF}
In this section we calculate the partition function as defined in (\ref{eq:PFformal}) for the system sketched in Fig.\ref{fig:Sys}. Our specification of the particle positions in the vectors $\textbf{X}$, $\textbf{U}$ and $\textbf{V}$ allows this to be done by following a simple scheme. First, let us consider the space between the leftmost wall, at $x_0 = 0$, and the first particle on the central layer at $x_1$. In this space there are $n_0^u$ and $n_0^v$ particles on the top and bottom layers respectively. These sets of particles are independent from each other and so their positions may be integrated over separately. This can be repeated for the particles between $x_1$ and $x_2$ and so forth until the final gap between $x_n$ and the right-hand wall at $x_{n+1}=L$. This yields an expression which depends on the co-ordinates $x_{k}$ of the particles on the central layer. These are finally integrated over to obtain $\mathcal{Z}(\textbf{N})$.
	\subsection{Top and Bottom Layers}
	\label{sec:TopBot}
	We start with the leftmost particle in the space between $x_0$ and $x_1$ at position $u_0(1)$. The wall to its left forces $w_0/2 < u_0(1)$. The next particle to the right is at position $u_0(2)$ and, due to the finite width of the particles, we must have $u_0(1) < u_0(2) - w_0$. Moving to the right, particle by particle, we can write similar inequalities for all of the ${u_0(j)}$
	\begin{equation}
		\begin{split}
			w_0/2 < \ &u_0(1) \ < u_0(2) - w_0,\\
			3 w_0/2 < \ &u_0(2) \ < u_0(3) - w_0,\\
			&\cdots \\
			\left(2 n_0 -1\right) w_0/2 < \ &u_0(n^u_0) \ < x_1 - w(h). 
		\end{split}
	\end{equation}
	The final inequality here \textit{defines} $w(h)$; the distance between the centres of a particle on the central layer and one on an outer layer when they just touch. This distance depends on the shape of the particle tip and is where the effect of particle geometry enters the problem. 
	
	Following the above inequalities, we can integrate over all allowed positions of these particles as follows
	\begin{equation}
		Z^u_0(x_1) = \int_{\left(2 n_0 -1\right) w_0/2}^{x_1-w(h)} du_0(n_0^u) \cdots \int_{3w_0/2}^{u_0(3)-w} du_0(2) \int_{w_0/2}^{u_0(2) - w_0} du_0(1).
	\end{equation}
	This is recognisably the partition function for a Tonks gas of $n_0^{u}$ particles with width $w_0$ occupying a length $x_1 + w_0/2 - w(h)$ \cite{Tonks1936TheSpheres}. The result is well known
 \begin{subequations}
	\begin{equation}
		\label{eq:Zu0}
		Z^u_0(x_1) = \frac{1}{n_0^u !}\left[x_1 - w(h) - \frac{2 n^u_0 -1}{2}w_0\right]^{n_0^u}.
	\end{equation}
	We can follow exactly the same procedure to integrate over the set of particles on the bottom layer between the wall and $x_1$. This yields
	\begin{equation}
		Z^v_0(x_1) = \frac{1}{n_0^v !}\left[x_1 - w(h) - \frac{2 n^v_0 -1}{2}w_0\right]^{n_0^v}.
	\end{equation}
	Note that, while we have implicitly assumed that $n_0^{u,v} \neq 0$ in these calculations, the result is correct when $n_0^{u,v} = 0$; unity. 
	
	For the sets of particles between $x_k$ and $x_{k+1}$, the calculation is essentially the same and the result of the integrals over all allowed $u_k(j)$ are
	\begin{equation}
		Z_k^u(x_{k+1}-x_k) = \frac{1}{n_k^u!}\bigg[x_{k+1} - x_k - \left(2w(h) + (n^u_k - 1)w_0\right)\bigg]^{n_k^u}.
	\end{equation}
	Similarly, for the bottom layer we find
	\begin{equation}
		Z_k^v(x_{k+1}-x_k) = \frac{1}{n_k^v!}\bigg[x_{k+1} - x_k - \left(2w(h) + (n^v_k - 1)w_0\right)\bigg]^{n_k^v}.
	\end{equation}
	
	The final sets of particles to consider are those between $x_n$ and the wall on the right. The integrals over their positions are done in much the same way, and we find 
	\begin{equation}
		Z_n^u(x_n) = \frac{1}{n_n^u!} \left[L - x_n - \left(w(h) + \frac{2 n_n^u - 1}{2} w_0 \right)\right]^{n_n^u},
	\end{equation}
	for the upper layer and 
	\begin{equation}
		\label{eq:Zvn}
		Z_n^v(x_n) = \frac{1}{n_n^v!} \left[L - x_n - \left(w(h) + \frac{2 n_n^v - 1}{2} w_0 \right)\right]^{n_n^v}
	\end{equation}
 \end{subequations}
	for the lower layer. 
	\subsection{Central Layer}
	\label{sec:Cent}
	Next we need to integrate over the allowed positions of the particles on the central layer. This as it was for the outer layers, but more attention must be paid to the integration limits. For simplicity, let us for now take $n_k^{u,v}\neq 0$ for each $k$. The case when there are some $n_k^{u,v} = 0$ shall be addressed at the end. We first consider the limits on $x_1$. This particle is, of course, constrained on the left by the wall. It is prevented from reaching the wall, however, by those particles on the top and bottom layers. Those on the top layer prevent $x_1$ being any smaller than $w(h) + (2 n_0^{u} -1) w_0/2$ and those on the bottom stop it reaching below $w(h) + (2 n_0^{v} -1) w_0/2$. Therefore, the smallest value that $x_1$ can reach is \textit{maximum} of these two. If we define the function
	\begin{equation}
		W_0(n) = w(h) + \frac{2 n - 1}{2} w_0, 
	\end{equation}
	then we have
	\begin{equation}
		W_0\left(\max_{u,v} \ n_0^{u,v}\right) < x_1,
	\end{equation}
	since $W_0(n)$ is monotonically increasing in its argument. 
	
	The upper limit for $x_1$ may be found similarly. It is prevented from reaching $x_2$ but the particles on the outer layers. Those on the top mean $x_1$ may not come within $2w(h) + (n_1^{u}-1)w_0$ of $x_2$ and those on the bottom make this distance $2w(h) + (n_1^{v}-1)w_0$. The upper limit on $x_1$ is then determined by the \textit{largest} of these two. We define the function 
	\begin{equation}
		W_1(n) = 2w(h) + (n-1)w_0,
	\end{equation}	
	so that this upper limit may be written 
	\begin{equation}
		x_1 < x_2 - W_1\left(\max_{u,v} n_1^{u,v}\right).
	\end{equation}
	This process is repeated for each $x_k$ so that the integrals of some function $A(x_1, \cdots, x_n)$, say, over all allowed $x_k$ may be computed as follows,
	\begin{equation}
		\label{eq:centint}
		\int_{\sum_{k=0}^{n-1} W_k\left(\max_{u,v} n_k^{u,v}\right)}^{L- W_n\left(\max_{u,v} n_n^{u,v}\right)} dx_n \cdots \int_{\sum_{k=0}^{j-1} W_k\left(\max_{u,v} n_k^{u,v}\right)}^{x_{j+1}- W_j\left(\max_{u,v} n_j^{u,v}\right)} dx_j \cdots \int_{W_0\left(\max_{u,v} n_0^{u,v}\right)}^{x_2- W_1\left(\max_{u,v} n_1^{u,v}\right)} dx_1 \ A(x_1, \cdots, x_n).  
	\end{equation}
	Note that we have defined the functions $W_k(n) = W_1(n)$ for the intermediate integrations and $W_n(n) = W_0(n)$ for the final integration. 
	\subsection{Putting it all together}
	\label{sec:alltog}
	We are now in a position to write out the partition function $\mathcal{Z}(\textbf{N})$ in full. Reiterating (\ref{eq:PFformal}), this is formally
	\begin{equation}
		\mathcal{Z}(\textbf{N}) = \int d\textbf{X} \int d\textbf{U}d\textbf{V} \ \Theta(\textbf{X},\textbf{U},\textbf{V}| \textbf{N}),
	\end{equation}
	The integrals over $\textbf{U}$ and $\textbf{V}$ are taken following section \ref{sec:TopBot}. This leaves the integrals over $x_k$ just as in (\ref{eq:centint}) but weighted by the various partition functions which result from the $\textbf{U}$ and $\textbf{V}$ integrals, see equations (\ref{eq:Zu0}-\ref{eq:Zvn}),
	\begin{equation}
		\label{eq:Zfull}
		\begin{split}
			&\mathcal{Z}(\textbf{N}) = \int_{\sum_{k=0}^{n-1} W_k\left(\max_{u,v} n_k^{u,v}\right)}^{L- W_n\left(\max_{u,v} n_n^{u,v}\right)} dx_n \cdots \int_{\sum_{k=0}^{j-1} W_k\left(\max_{u,v} n_k^{u,v}\right)}^{x_{j+1}- W_j\left(\max_{u,v} n_j^{u,v}\right)} dx_j \cdots \int_{W_0\left(\max_{u,v} n_0^{u,v}\right)}^{x_2- W_1\left(\max_{u,v} n_1^{u,v}\right)} dx_1 \\
			&\times Z_n^{u}(x_n)Z_n^{v}(x_n) Z_{n-1}^{u}(x_n - x_{n-1})Z_{n-1}^{v}(x_n - x_{n-1}) \cdots Z_1^{u}(x_2-x_1)Z_1^{v}(x_2-x_1)Z_0^{u}(x_1)Z_0^{v}(x_1).
		\end{split}
	\end{equation}
	This expression appears formidable, but it has a relatively simple structure which is revealed after some manipulations. First, we write the $Z_k^{u,v}$ out in full using the functions $W_k(n)$:
	\begin{subequations}
		\begin{equation}
			\label{eq:Zu0W}
			Z^u_0(x_1) = \frac{1}{n_0^u !}\left[x_1 -W_0(n_0^{u})\right]^{n_0^u}, \ \ \text{and} \ \	Z^v_0(x_1) = \frac{1}{n_0^v !}\left[x_1 -W_0(n_0^{v})\right]^{n_0^v}.
		\end{equation}
		\begin{equation}
			Z_k^u(x_{k+1}-x_k) = \frac{1}{n_k^u!}\bigg[x_{k+1} - x_k - W_k(n_k^{u})\bigg]^{n_k^u}, \ \ \text{and} \ \ Z_k^v(x_{k+1}-x_k) = \frac{1}{n_k^v!}\bigg[x_{k+1} - x_k - W_k(n_k^{v})\bigg]^{n_k^v}.
		\end{equation}
		\begin{equation}
  \label{eq:ZvnW}
			Z_n^u(x_n) = \frac{1}{n_n^u!} \left[L - x_n - W_n(n_n^{u})\right]^{n_n^u}, \ \ \text{and} \ \ Z_n^v(x_n) = \frac{1}{n_n^v!} \left[L - x_n - W_n(n_n^{v})\right]^{n_n^v}.
		\end{equation}
	\end{subequations}
	The next step is to transform from $x_k$ to the variables $y_k$ defined through
	\begin{equation}
		y_k = x_k - \sum_{j=0}^{k-1}W_k\left(\max_{u,v} n_k^{u,v}\right).
	\end{equation}
	This transformation is convenient because it simplifies the integration limits in (\ref{eq:Zfull}); the lower limit on any $y_k$ is zero and its upper limit is $y_{k+1}$, with the upper limit on the final integration over $y_n$ being $L - \sum_{k=0}^{n}W\left(\max_{u,v} n_k^{u,v}\right)$. In terms of the new variables $y_k$ the $Z_k^{u,v}$ may also be simplified somewhat. If we focus on one particular $k$ we have
	\begin{subequations}
		\begin{equation}
			Z_{k}^{u}(y_{k+1}-y_k) = \frac{1}{n_k^u!}\left[y_{k+1} - y_k +\left(W_k\left(\max_{u,v} n_k^{u,v}\right) - W_k(n_k^{u})\right)\right]^{n_k^u},
		\end{equation} 
		and
		\begin{equation}
			Z_{k}^{v}(y_{k+1}-y_k) = \frac{1}{n_k^v!}\left[y_{k+1} - y_k +\left(W_k\left(\max_{u,v} n_k^{u,v}\right) - W_k(n_k^{v})\right)\right]^{n_k^v}.
		\end{equation} 
	\end{subequations}
	Let us define, 
	\begin{equation}
		\Delta_k^{u,v} = W_k\left(\max_{u,v} n_k^{u,v}\right) - W_k(n_k^{v}) = \left(\max_{u,v} n_k^{u,v} - n_k^{v}\right) w_0.
	\end{equation}
	It is evident that one of $\Delta_k^{u}$ or $\Delta_k^{v}$ will be zero corresponding to whichever of $n_k^{u}$ or $n_k^v$ is the largest. For each $k$, let us call the largest of these two $n_k$ and the smallest $m_k$. We shall also call the non-zero $\Delta_k^{u,v}$ simply $\Delta_k$ and note that it is positive. The partition function is then written in full as
	\begin{equation}
		\label{eq:Zfull2}
		\begin{split}
			\mathcal{Z}(\textbf{N}) =& \int_{0}^{L-\sum_{k=0}^{n}W_k(n_k)}dy_n \int_{0}^{y_n}dy_{n-1}\cdots \int_{0}^{y_{j+1}} dy_j \cdots \int_{0}^{y_2} dy_1 \\ &\times\prod_{k=0}^{n}\frac{1}{n_k! m_k!} \left(L-\sum_{k=0}^{n}W_k(n_k) - y_n\right)^{n_n}\left(L-\sum_{k=0}^{n}W_k(n_k) - y_n+\Delta_n\right)^{m_n} \\
			&\times \left(y_n - y_{n-1}\right)^{n_{n-1}}\left(y_n - y_{n-1} + \Delta_{n-1}\right)^{m_{n-1}}\cdots y_1^{n_0} \left(y_1 +\Delta_0\right)^{m_0}.	
		\end{split}
	\end{equation}
	Hence the partition function takes the form of an $n$-fold convolution, as is typical for one-dimensional systems, of which ours is a more elaborate example \cite{Lieb1966}. This structure allows us to employ the Laplace transform to make progress. We are working in the canonical ensemble, so the Laplace transform will actually give us the partition function in the isobaric (constant pressure) ensemble directly \cite{kardar_2007}. However, as we shall see, its form is very unwieldy and we have not found it to be of much use. Instead, a natural approximation shall reveal itself as the way forward in the canonical ensemble. Before discussing this, let us return to the case when there are some $m_k$ and $n_k$ that are zero. When this is so, some of the partition functions $Z_k^{u,v}$ will be unity. This case is already handled by the formulae (\ref{eq:Zu0W}-\ref{eq:ZvnW}). If only one of $n_k$ or $m_k$ is zero (and by definition it must be $m_k$) nothing in the calculation needs to change. 
	
	The case when both are zero requires more thought. This is because the lower limit of the $x_k$ integral will be different. This is seen most easily by considering $x_1$. When there are no particles on the outer layers between the wall and $x_1$, then the smallest $x_1$ can be is $w_0/2$. This is \textit{not} the result returned by $W_0(n=0)$. Therefore, in order to take into account the case when both $n_k = m_k = 0$, we simply need to alter the definitions of the functions $W_k(n)$ so that when $n=0$ it returns the required value. The corrected expressions are
	\begin{subequations}
		\label{eq:Wks}
		\begin{equation}
			W_0(n_0) = w(h) + \frac{2n_0-1}{2} w_0 + \left(w_0 - w(h)\right)\delta(n_0)
		\end{equation}
		\begin{equation}
		W_k(n_k) = 2w(h) + (n_k-1)w_0 + 2\left(w_0 - w(h)\right)\delta(n_k)
	\end{equation}
\begin{equation}
	W_n(n_n) = w(h) + \frac{2n_n-1}{2} w_0 + \left(w_0 - w(h)\right)\delta(n_n),
\end{equation}
	\end{subequations} 
 where we have used the Kronecker delta; $\delta(n) = 1$ if $n=0$ and is zero otherwise. With these changes the expression (\ref{eq:Zfull2}) for the partition function works for any values of $n_k^{u,v}$ despite the fact that it was derived assuming they were all non-zero. 
	\subsection{Laplace Transform}
	\label{sec:LapTran}
	Before we proceed to apply the Laplace transform to (\ref{eq:Zfull2}), it is convenient to re-scale the $y_k$ to $z_k = y_k / L$. This yields
	\begin{equation}
			\label{eq:Zfull3}
			\begin{split}
				&\mathcal{Z}(\textbf{N},\xi) = \prod_{k=0}^{n}\frac{L^N}{n_k! m_k!} \int_{0}^{\xi}dz_n \int_{0}^{z_n}dy_{z-1}\cdots \int_{0}^{z_2} dz_1 \left(\xi - z_n\right)^{n_n}\left(\xi - z_n+\Delta_n/L\right)^{m_n}\\
				&\times \left(z_n - z_{n-1}\right)^{n_{n-1}}\left(z_n - z_{n-1} + \Delta_{n-1}/L\right)^{m_{n-1}}\cdots z_1^{n_0} \left(z_1 +\Delta_0/L\right)^{m_0}.	
			\end{split}
		\end{equation}
	where we have introduced $\xi = 1 - \sum_k W_k(n_k) /L$. Let us now define the Laplace transform of $\mathcal{Z}(\textbf{N},\xi)$ as
	\begin{equation}
		\mathcal{Q}(\textbf{N},\zeta) = \int_{0}^{\infty} d \xi \ e^{- \zeta \xi} \mathcal{Z}(\textbf{N},\xi).
	\end{equation}
	If we define the $n$-fold integral in (\ref{eq:Zfull3}) as $F_n(\xi)$, so that 
	\begin{equation}
		\mathcal{Z}(\textbf{N},\xi) = \prod_{k=0}^{n}\frac{L^N}{n_k! m_k!} F_n(\xi),
	\end{equation}
we can extract a recursive definition for $F_n$
	\begin{equation}
	\label{eq:FnIntRec}
	F_n(\xi) = \int_0^{\xi} dy \ (\xi - y)^{n_n} (\xi - y + \Delta_n/L)^{m_n} F_{n-1}(y).
\end{equation} 
Or, if we write the Laplace transform of $F_n$ as $\mathcal{F}_n$,
	\begin{equation}
		\label{eq:Fmult}
	\mathcal{F}_n(\zeta) = (\Delta_n/L)^{n_n + m_n + 1}  n_n! \mathcal{U}(n_n + 1, 2 + m_n + n_n, \Delta_n \zeta /L) \mathcal{F}_{n-1}(\zeta).
\end{equation}
Here $\mathcal{U}(a , b , x)$ is the solution to Kummer's equation which is singular at the origin (the confluent hypergeometric function of the second kind) \footnote{see Chap. 13 of \cite{Abramowitz1964HandbookTables}} and we have made use of its integral representation 
\begin{equation}
	\mathcal{U}(a,b,x) = \int_{0}^{\infty} dt \ e^{-x t} t^{a-1} (t+1)^{b-a-1}.
\end{equation}
With the multiplicative relation (\ref{eq:Fmult}) we find
	\begin{equation}
	\label{eq:Ziso}
	\begin{split}
		&\mathcal{Q}(\textbf{N},\zeta) = L^N \prod_{k=0}^{n} \frac{\left((n_k - m_k)w_0 /L\right)^{n_k + m_k + 1}}{m_k!} \mathcal{U}\left(n_k + 1, 2 + m_k + n_k, (n_k-m_k)\frac{w_0}{L} \zeta\right).
	\end{split}
\end{equation}
It is well known that the partition function in the Isobaric (constant pressure) ensemble is related to its canonical cousin by the Laplace transform \cite{kardar_2007}. Therefore, this expression represents an exact solution for that function. While it is possible to use this to find expressions for thermodynamic quantities such as the chemical potential and hence equations of state, the complexity of (\ref{eq:Ziso}) means they are not so insightful. Fortunately, we may proceed by means of a natural approximation.
\subsection{Approximation}
\label{sec:Appox}
The essential difficulty with (\ref{eq:Ziso}) is the appearence of Kummer's function $\mathcal{U}$. However, notice that its argument contains the combination $(n_k - m_k) w_0 /L$. In the thermodynamic limit that $L \to \infty$ we can expect this to be very small. This is because, in order for it to be of order unity, we must have $n_k - m_k \sim N$. Given that $m_k \leq n_k$ this would require $n_k \sim N$ or, in other words, almost all of the particles to be on one layer. Intuitively, this situation is extremely unlikely given that each particle would be afforded more space by moving to a different layer, not to mention the entropy gained by the system in re-arranging the particles between the layers. Under the assumption that $(n_k - m_k) w_0 /L \ll 1$, we may expand each term of the product in $\mathcal{Q}$ in that limit
\begin{equation}
	\mathcal{Q}(\textbf{N},\zeta) \approx L^N \prod_{k=0}^{n}\frac{(n_k + m_k)!}{n_k! m_k!} \zeta^{-1-n_k-m_k}\left[1 + \frac{w_0}{L}\frac{m_k(n_k - m_k)}{n_k + m_k} \zeta + \mathcal{O}\left((w_0/L)^2\right)\right].
\end{equation}
Then using $\prod_k f_k(1 + \delta_k ) \approx \prod_k f_k (1 + \sum_l \delta_l)$, we find to leading order
\begin{equation}
	\mathcal{Q}(\textbf{N},\zeta) \approx  L^N \left[\prod_{k=0}^{n}\frac{(n_k + m_k)!}{n_k! m_k!}\right] \zeta^{-(N+1)}\left(1 + \frac{w_0}{L} \zeta\sum_{k=0}^{n}\frac{m_k(n_k - m_k)}{n_k + m_k} + \cdots \right).
\end{equation}
We can now invert the Laplace transform term by term
\begin{equation}
	\mathcal{Z}(\textbf{N},\xi) = \frac{(L \xi)^N}{N!} \left[\prod_{k=0}^{n}\frac{(n_k + m_k)!}{n_k! m_k!}\right] \left(1 + \frac{N w_0}{L \xi}\sum_{k=0}^{n}\frac{m_k(n_k - m_k)}{n_k + m_k} + \cdots\right).
\end{equation}
Finally, if we recall the definition $\xi = 1 - \sum_k W_k(n_k)/L$ and additionally define the ``length fraction'' $\varphi = N w_0 / L \leq 2$, we obtain a usable expression for the canonical partition function
\begin{equation}
	\begin{split}
		\label{eq:PFFinalcomplex}
\mathcal{Z}(\textbf{N}) =& \frac{1}{N!}\prod_{k=0}^{n}\frac{(n_k + m_k)!}{n_k! m_k!}\left(L - \sum_{k=0}^{n}W_k(n_k)\right)^N \\
&\times\left[1 + \frac{\varphi}{1 - \sum_{k=0}^{n}W_k(n_k)/L}\sum_{k=0}^{n}\frac{m_k(n_k - m_k)}{n_k + m_k} + \cdots\right].
	\end{split}
\end{equation}
An important feature of this is that the square bracket \textit{is not} something to the power $N$. We consider the system in the thermodynamic limit $N \to \infty$ and so this factor will only give sub-dominant terms. Therefore, from here onwards, it shall be dropped leaving a relatively simple expression for the partition function
\begin{equation}
		\label{eq:PFFinal}
		\mathcal{Z}(\textbf{N}) \approx \frac{1}{N!}\prod_{k=0}^{n}\frac{(n_k + m_k)!}{n_k! m_k!}\left(L - \sum_{k=0}^{n}W_k(n_k)\right)^N 
\end{equation}
The final step to make is to simplify the summation $\sum_{k} W_k(n_k)$. Using (\ref{eq:Wks}) this can be written 
\begin{equation}
	\label{eq:Wcomp}
	\begin{split}
		\sum_{k=0}^{n} W_k(n_k) &= \left(2w(h)-w_0\right) n + \sum_{k=0}^{n}w_0 n_k + 2(w(h)-w_0) \sum_{k=0}^{n} \delta(n_k) \\
		&- \left(w(h)-w_0\right)\left(\delta(n_0) + \delta(n_n)\right).
	\end{split}
\end{equation}
The final term originates from the final particles on the outer layers, sandwiched between a particle on the central layer and a wall. These terms essentially quantify the fact that the wall always excludes a length $w_0/2$ to any particle, independent of the layer it occupies. Retaining these boundary terms adds unnecesary complication and it may be shown that they only introduce sub-dominant corrections as $N \to \infty$. Therefore, they shall also be dropped, leaving 
\begin{equation}
	\label{eq:W}
	\begin{split}
		\sum_{k=0}^{n} W_k(n_k) &\approx \left(2w(h)-w_0\right) n + \sum_{k=0}^{n}w_0 n_k + 2(w(h)-w_0) \sum_{k=0}^{n} \delta(n_k). 
	\end{split}
\end{equation}
\section{Smectic vs. Nematic}
\label{sec:SmectTran}
Now that we have a usable expression for the partition function $\mathcal{Z}(\textbf{N})$ we can follow the scheme outlined in section \ref{sec:Model} to investigate the smectic transition. The first step is to compute the free energy of the smectic state. 
\subsection{Smectic Free Energy}
\label{sec:SmecFreeEn}
For the smectic state, $n_k = m_k$, so the approximation introduced in the previous section is not required to invert the Laplace transform of (\ref{eq:Ziso}). Recalling the definitions (\ref{eq:NSm}) and (\ref{eq:ZSm}) we find the exact solution
\begin{equation}
	\mathcal{Z}_{\text{Sm}} = \frac{1}{N!} \binom{N}{N/2} \big(L - N w_0/2\big)^{N}.
\end{equation}
This result could have been anticipated since it is the product of the partition functions of two separate Tonks gasses each with $N/2$ particles of width $w_0$ multiplied by the number of ways of distributing the particles between the two layers. Taking the logarithm leads to the free energy which, in the limit $N \to \infty$, may be written 
\begin{equation}
	\beta F_{\text{Sm}} = - N \left[\log (1 - \varphi/2) + \log 2 + 1 - \log \rho\right],
\end{equation}
where the number density $\rho = N/L$ and length fraction $\varphi = w_0 \rho$ have been used.
\subsection{Nematic Free Energy}
\label{sec:NemFreeEn}
The nematic free energy is much harder sought than that of the smectic. Our first task is to determine the weighting as described in section \ref{sec:Model}. Let us focus on, $\mathcal{P}(\textbf{N})$, the proportion of nematic states with a given $\textbf{N}$. In the nematic, by definition, the number of particles on the central layer is fixed at $N/3$. This means there are $2N/3$ left to share between the outer layers. These are to be distributed among the $N/3 + 1$ gaps between the central particles such that there is one group of $n_0$ and one of $m_0$ in the first, one of $n_1$ and one of $m_1$ in the second, and so on. The number of ways of doing this is given by the multinomial co-efficient
\begin{equation}
	\frac{(2N/3)!}{\prod_{k=0}^{N/3} n_k! m_k!}.
\end{equation}
To obtain the proportion $\mathcal{P}$, this must be divided by its sum over all sets of $n_k$ and $m_k$ whose total sum is $2N/3$. Using the multinomial theorem, this is $(2N/3)^{2N/3}$ and hence,
\begin{equation}
	\mathcal{P}(\textbf{N})= \frac{1}{ (2N/3)^{2N/3}}\frac{(2N/3)!}{\prod_{k=0}^{N/3} n_k! m_k!}.
\end{equation}
Next we require the total number of nematic states, $\mathcal{N}$. Again we make use of the fact that in the nematic the occupancy of the central layer is fixed at $N/3$. Then any nematic state is a way of dividing the $2N/3$ particles on the outer layers into two equal parts. The number of ways of doing this is
\begin{equation}
	\mathcal{N} = \binom{2N/3}{N/3}.
\end{equation}
This gives the appropriate weighting to compute the nematic partition function in (\ref{eq:ZNemForm}). The final puzzle piece is to constrain the sums over $n_k$ and $m_k$ so that the two outer layers have equal occupancy. To do this, we introduce the variables $\sigma_k$. These specify which of the outer layers holds the most particles between the $k^{\text{th}}$ and $(k+1)^{\text{st}}$ central particles, i.e. does $n_k$ belong to the upper or lower layer. These take the values $\pm 1$ depending on the situation, namely
\begin{equation}
	\label{eq:SigDef}
	\sigma_{k} = 
		\begin{cases}
			+1, & \text{if $n_k$ on upper layer}\\
			-1, & \text{if $n_k$ on lower layer}
		\end{cases}
\end{equation}
The introduction of the variables $\sigma_k$ is important; while the partition function $\mathcal{Z}(\textbf{N})$ does not depend on the $\sigma_k$ the constraints (\ref{eq:NemConst}) which enforce the nematic state do. Using our definition (\ref{eq:SigDef}), we can re-write (\ref{eq:NemConst}) as
\begin{subequations}
	\label{eq:NemConstSig}
	\begin{equation}
		\sum_{k=0}^{N/3} n_k^{u} = \frac{1}{2}\sum_{k=0}^{N/3} (\sigma_k +1)n_k - (\sigma_k -1)m_k = \frac{N}{3}
	\end{equation}
\begin{equation}
	\sum_{k=0}^{N/3} n_k^{v} = \frac{1}{2}\sum_{k=0}^{N/3} (\sigma_k +1)m_k - (\sigma_k -1)n_k = \frac{N}{3}
\end{equation}
\end{subequations}
All of these pieces let us write a concrete expression for the nematic partition function
\begin{equation}
		\mathcal{Z}_{\text{N}} = \binom{2N/3}{N/3} \frac{(2N/3)!}{(2N/3)^{2N/3}}\frac{1}{N!} \prod_{k=0}^{N/3}\sum_{\substack{\sigma_{k}= \pm 1 \\ n_k \geq 0 \\ m_k \leq n_k}}^{*} \frac{(n_k +m_k)!}{(n_k !)^2 (m_k !)^2} \left(L - \sum_{k=0}^{N/3}W_k(n_k)\right)^N
\end{equation}
The star on the summation above means that the $\sigma_k$, $n_k$ and $m_k$ are subject to the constraints (\ref{eq:NemConstSig}), these variables are also restricted by the conditions underneath the summation symbol. Let us use (\ref{eq:W}) to manipulate the final bracket. We start with
\begin{equation}
	\left(L - \sum_{k=0}^{N/3} W_k(n_k)\right)^N \approx \left(L - \frac{N}{3}\left(2 w(h) - w_0\right) - \sum_{k=0}^{N/3} w_0 n_k - \sum_{k=0}^{N/3} 2 \left(w(h) - w_0\right) \delta(n_k)\right)^N.
\end{equation}
We pull the first two terms out of the bracket on the right hand side, and re-write everything in terms of the length fraction, $\varphi$, and the relative tip width, $x = w(h)/w_0 \leq 1$, to obtain
\begin{equation}
L^N\left(1 - \frac{\varphi}{3} (2x-1)\right)^N \left(1- \frac{1}{N}\sum_{k=0}^{N/3} \frac{\varphi \ n_k}{1- \varphi (2x-1)/3} + \frac{2\varphi (1-x) \delta(n_k)}{1 - \varphi (2x-1)/3}\right)^N.
\end{equation}  
Defining 
\begin{equation}
	\alpha = \frac{\varphi}{1-\varphi(2x-1)/3},
\end{equation}
and noticing that the final bracket is the definition of the exponential in the limit $N \to \infty$, we have
\begin{equation}
	\left(L - \sum_{k=0}^{N/3} W_k(n_k)\right)^N \approx L^N \left(1 - \frac{\varphi}{3} (2x-1)\right)^N \exp\left(- \sum_{k=0}^{N/3} \alpha n_k + 2 \alpha (1-x) \delta(n_k) \right).
\end{equation}
This gives the expression for the nematic partition function in a usable form,
\begin{equation}
	\begin{split}
	\mathcal{Z}_{\text{N}} &= \binom{2N/3}{N/3} \frac{(2N/3)!}{(2N/3)^{2N/3}}\frac{L^N}{N!} \left(1 - \frac{\varphi}{3} (2x-1)\right)^N \\ & \times \prod_{k=0}^{N/3}\sum_{\substack{\sigma_{k}= \pm 1 \\ n_k \geq 0 \\ m_k \leq n_k}}^{*} \frac{(n_k +m_k)!}{(n_k !)^2 (m_k !)^2} \exp \big[- \alpha n_k - 2 \alpha (1-x) \delta(n_k)\big].
	\end{split}
\end{equation}
The first line of the above equation is easy to handle but the second is more challenging. Let us define
\begin{equation}
	\Upsilon(N) \equiv\prod_{k=0}^{N/3}\sum_{\substack{\sigma_{k}= \pm 1 \\ n_k \geq 0 \\ m_k \leq n_k}}^{*} \frac{(n_k +m_k)!}{(n_k !)^2 (m_k !)^2} \exp \big[- \alpha n_k - 2 \alpha (1-x) \delta(n_k)\big].
\end{equation}
The large $N$ behaviour of this function controls the nematic free energy and is our target but the conditioning of the sum makes this difficult. However, this may be overcome by a well known trick where the constraints are imposed by delta functions. Given the complexity of the above expression, it is useful to introduce this in a simpler setting. This is done in appendix \ref{app:Trick} and we direct the reader there for an outline of the approach.

\begin{figure}\includegraphics[width=16cm]{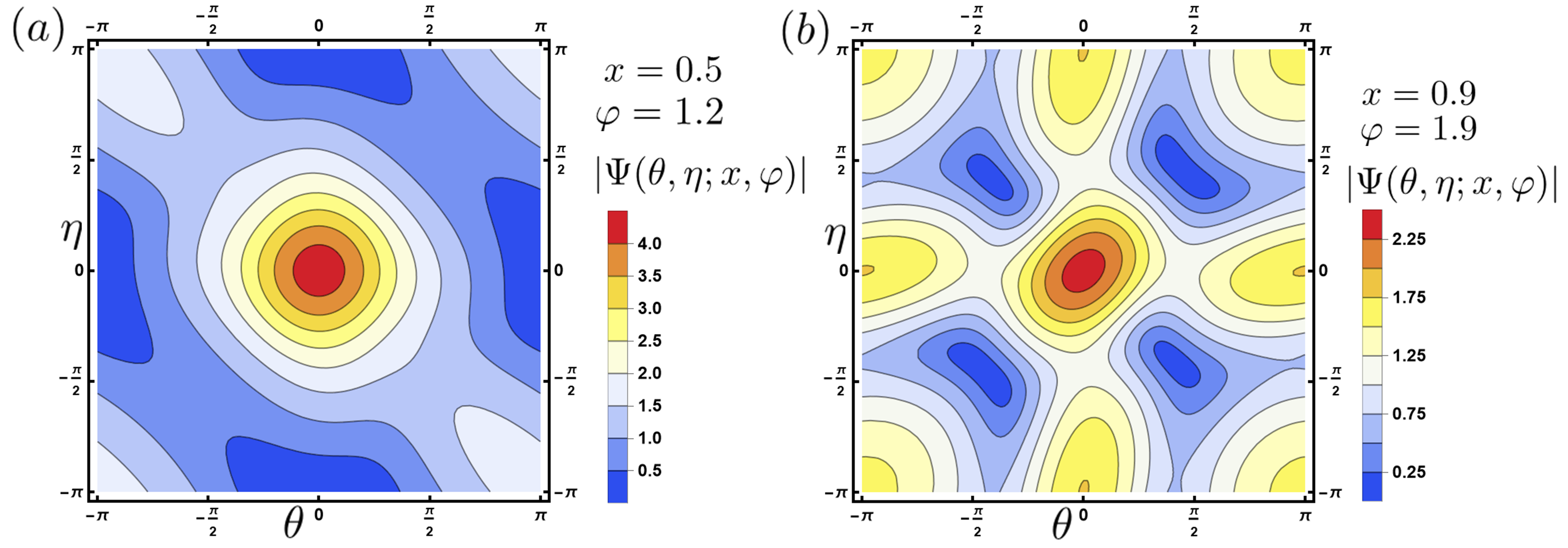}
		\caption{\label{fig:PhiPlot} Contour plots of $|\Psi(\theta,\eta;x,\varphi)|$. To compute $\Psi$, we truncated the sum in (\ref{eq:PhiBessel}) after 10 terms, which makes an imperceptible difference ($\sim 10^{-6}\%$) from using 20 terms. For each panel we have chosen characteristic values of $x$ and $\varphi$. Panel (a) shows a thin tipped and low density case; $x = 0.5$ and $\varphi = 1.2$. Panel (b) shows a wider tipped and high density case;  $x = 0.5$ and $\varphi = 1.2$. In each case, it is clear that the maximum of $|\Psi|$ occurs when $\theta = \eta = 0$. }
	\end{figure}

The first step is to impose (\ref{eq:NemConstSig}) by two delta functions, which may be written as the following fourier integrals 
\begin{subequations}
	 \begin{equation}
	 	\begin{split}
	 		 &\delta\left(\frac{1}{2}\sum_{k=0}^{N/3} (\sigma_k +1)n_k - (\sigma_k -1)m_k-N/3\right)\\ 
	 		 &= \int_{-\pi}^{\pi} \frac{d \theta}{2\pi} e^{- i N \theta/3} \exp\left[i\sum_{k=0}^{N/3} \frac{1}{2}(\sigma_k +1) n_k \theta \right] \exp\left[-i\sum_{k=0}^{N/3} \frac{1}{2}(\sigma_k -1)m_k\theta  \right] 
	 	\end{split}
	 \end{equation}
 	 \begin{equation}
 	\begin{split}
 		&\delta\left(\frac{1}{2}\sum_{k=0}^{N/3} (\sigma_k +1)m_k - (\sigma_k -1)n_k-N/3\right)\\ 
 		&= \int_{-\pi}^{\pi} \frac{d \eta}{2\pi} e^{- i N \eta/3} \exp\left[i\sum_{k=0}^{N/3} \frac{1}{2}(\sigma_k +1) m_k \eta \right] \exp\left[-i\sum_{k=0}^{N/3} \frac{1}{2}(\sigma_k -1)n_k\eta  \right]
 	\end{split}
 \end{equation}
\end{subequations}
These allow $\Upsilon$ to be written in terms of separate un-constrained sums on $\sigma_k$, $n_k$ and $m_k$,
\begin{equation}
	\label{eq:PF1}
	\begin{split}
	\Upsilon(N) &=  \int_{-\pi}^{\pi} \frac{d \theta d\eta}{(2 \pi)^2} e^{- iN ( \theta + \eta)/3}\prod_{k=0}^{N/3} \sum_{\sigma_{k} = \pm 1}\sum_{n_k = 0}^{N/3} \sum_{m_k =0}^{n_k} \frac{(n_k +m_k)!}{(n_k !)^2 (m_k !)^2}  e^{-2 \alpha (1- x)\delta(n_k)} e^{(i \beta_k - \alpha)n_k} e^{i \gamma_k m_k}.
	\end{split}
\end{equation}
Here we have introduced 
\begin{subequations}
	\label{eq:BetaGammaDefs}
	\begin{equation}
		\beta_k = \frac{1}{2}(\sigma_k + 1) \theta - \frac{1}{2}(\sigma_k - 1) \eta,
	\end{equation}
and
	\begin{equation}
	\gamma_k = \frac{1}{2}(\sigma_k + 1) \eta - \frac{1}{2}(\sigma_k - 1) \theta.
\end{equation}
\end{subequations}
We can remove the delta function in the exponent of (\ref{eq:PF1}) as follows, 
\begin{equation}
	\label{eq:PF2}
	\begin{split}
		\Upsilon(N) &=  \int_{-\pi}^{\pi} \frac{d \theta d\eta}{(2 \pi)^2}e^{- iN ( \theta + \eta)/3}\prod_{k=0}^{N/3} \sum_{\sigma_{k} = \pm 1}\left[e^{-2 \alpha (1-x)} - 1 + \sum_{n_k = 0}^{N/3} \sum_{m_k =0}^{n_k} \frac{(n_k +m_k)!}{(n_k !)^2 (m_k !)^2} e^{(i \beta_k - \alpha)n_k} e^{i \gamma_k m_k}\right].
	\end{split}
\end{equation}
We shall define
\begin{equation}
	\label{eq:PhiDef}
	\Phi(y,z) = \sum_{n = 0}^{N/3} \sum_{m =0}^{n} \frac{(n +m)!}{(n !)^2 (m !)^2} y^n z^m,
\end{equation}
so that we can write (\ref{eq:PF2}) more compactly
\begin{equation}
	\Upsilon(N) = \int_{-\pi}^{\pi} \frac{d \theta d\eta}{(2 \pi)^2}\prod_{k=0}^{N/3} \sum_{\sigma_{k} = \pm 1}\left[e^{-2 \alpha (1-x)} - 1 + \Phi(e^{i \beta_k - \alpha} , e^{i \gamma_k}) \right].
\end{equation}
From (\ref{eq:BetaGammaDefs}) we note that
\begin{equation}
	\beta_k = \begin{cases}
		\theta, \ \ \ \text{when} \  \sigma_k = 1 \\
		\eta, \ \ \ \text{when} \  \sigma_k = -1 
	\end{cases} \ \ \text{and} \ \ \gamma_k = \begin{cases}
	\eta, \ \ \ \text{when} \  \sigma_k = 1 \\
	\theta, \ \ \ \text{when} \  \sigma_k = -1 
\end{cases}.
\end{equation}
Hence, to calculate the partition function, we must evaluate the following double integral,
\begin{equation}
	\label{eq:PF3}
	\begin{split}
		\Upsilon(N) &=  \int_{-\pi}^{\pi} \frac{d \theta d\eta}{(2 \pi)^2} \exp\left(i \frac{N}{3}\left(\theta + \eta \right) + \frac{N}{3} \log\left[2\left(e^{-2 \alpha (1-x)} - 1\right)+\Phi(e^{i \theta - \alpha}, e^{i \eta})+\Phi(e^{i \eta - \alpha}, e^{i \theta})\right] \right) \\
  &\equiv \int_{-\pi}^{\pi} \frac{d \theta d\eta}{(2 \pi)^2} \exp\left(i \frac{N}{3}\left(\theta + \eta \right) + \frac{N}{3} \log \Psi(\theta,\eta; x,\varphi) \right).
   \end{split}
\end{equation}
Here we have defined $\Psi$ as the function appearing in the logarithm above, note that this function depends on the parameters $x$ and $\varphi$ through $\alpha$. This integral is to be taken in the limit $N \to \infty$ so, as may be shown by steepest descent \cite{Copson1965,Bender1978}, the dominant term will come from the maximum value of the real part of the function in the exponent. This is where the absolute value of $\Psi$ is maximised. In order to determine this, we must understand the function $\Phi$. This is made easier by noting that since we are working in the limit $N \to \infty$ we can take the upper limit on the sum over $n$ in (\ref{eq:PhiDef}) to infinity
\begin{equation}
	\label{eq:PhiDefInf}
		\Phi(y,z) \approx \sum_{n = 0}^{\infty} \sum_{m =0}^{n} \frac{(n +m)!}{(n !)^2 (m !)^2} y^n z^m.
\end{equation}
We are not able to cast this sum in a simple closed form. However, it may be more neatly written as an infinite series of modified Bessel functions as follows
\begin{equation}
	\label{eq:PhiBessel}
	\Phi(y,z)  = I_0(2 \sqrt{y z}) \left[e^{y+z} - \sum_{n=1}^{\infty} \left(z/y\right)^{n/2} I_n (2 \sqrt{y z}) \right],
\end{equation}
where $I_n(x)$ is the $n^{\text{th}}$ order modified Bessel function of the first kind. The manipulations to obtain this form are somewhat involved and are given in appendix \ref{app:ExpGenL}. The advantages of this series over (\ref{eq:PhiDefInf}) are its efficiency, fewer terms of the series in (\ref{eq:PhiBessel}) are required to obtain the same accuracy as (\ref{eq:PhiDefInf}), and the locations of its maximum values may be identified easily. This is done analytically in appendix \ref{app:PhiMax}, but it is sufficient to plot $|\Psi(\theta,\eta;x,\varphi)|$ numerically, as is done in Fig.(\ref{fig:PhiPlot}). This shows that its maximum occurs when $\theta = \eta = 0$, and hence that is where the real part of the integrand in (\ref{eq:PF3}) is maximised. Steepest descent or saddle point arguments then show that the dominant contribution to as $\Upsilon(N)$ as $N \to \infty$ is 
\begin{equation}
\Upsilon(N) \sim 2^{N/3}\ \left(e^{-2 \alpha (1-x)} - 1+\Phi(e^{- \alpha}, 1)\right)^{N/3}.
\end{equation}
At last, we have an expression for the nematic partition function 
\begin{equation}
	\begin{split}
		\mathcal{Z}_{\text{N}} \sim \frac{L^N}{N!} \binom{2N/3}{N/3} \frac{2^{N/3}(2N/3)!}{(2N/3)^{2N/3}}\left(1 - \frac{\varphi}{3} (2x-1)\right)^N \left(e^{-2 \alpha (1-x)} - 1+\Phi(e^{- \alpha}, 1)\right)^{N/3},
	\end{split}
\end{equation}
and taking its logarithm yields the free energy
\begin{equation}
	\beta F_{N}	= - N \left[\log\left(1 - \frac{\varphi}{3} (2x-1)\right)+ \frac{1}{3} \log f(\varphi,x) - \frac{2}{3} + \log 2  + 1 - \log \rho \right],
\end{equation}
where $f(\varphi,x) = e^{-2 \alpha (1-x)} - 1+\Phi(e^{- \alpha}, 1)$. Note that this is extensive (linear in $N$), as required
\subsection{Phase Diagram}
\label{sec:PhDiag}
In order to understand when the system prefers a smectic phase we need to determine where, as a function of $\varphi$ and $x$, we have $\beta F_{\text{Sm}} < \beta F_{\text{N}}$. In other words
\begin{equation}
	\label{eq:SmecIneq}
	 \left(\frac{1-(2x-1)\varphi /3}{1 -\varphi/2}\right)^3  f(\varphi,x) < e^2.
\end{equation}
Before analysing this inequality, which must be done numerically, it is worth comparing it to the condition for smecticity (\ref{eq:CrudeInEq}) obtained from the heuristic treatment in section \ref{sec:Inform}. Both the left hand side of the above and (\ref{eq:CrudeInEq}) depend on the density and geometry of the particles. They each originate from the entropic penalty when a particle is introduced to the central layer. The right hand side is a constant measuring the increased number of ways of realising the system when all three layers filled than when the central layer is empty. Therefore, (\ref{eq:SmecIneq}) articulates the same physics as the crude model, but the effects are calculated in detail, leading to more complicated functional forms. 
	\begin{figure}\includegraphics[width=8cm]{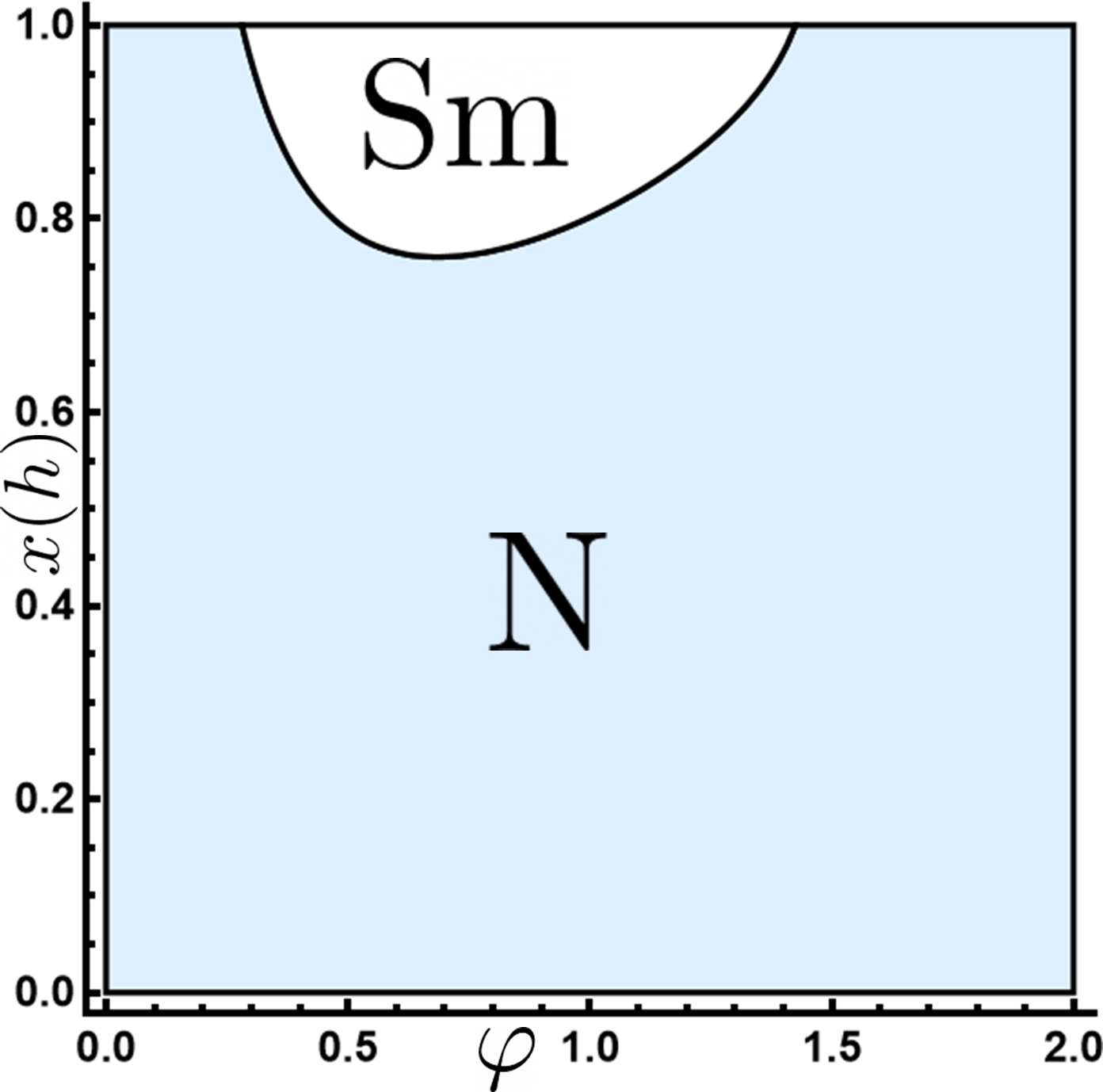}
		\caption{\label{fig:PDSm} The phase diagram for the system plotted with the length fraction of the particles $\varphi = N w_0 /L$ on the $x$-axis and the reduced tip width $x(h)= w(h)/w_0$ on the $y$-axis. The blue shaded region indicates where the nematic is more stable than the smectic $\beta F_{\text{N}} < \beta F_{\text{Sm}}$, and the unshaded region marks where the smectic is more stable. Both the tip width and the particle density need to be large enough for the smectic phase to be preferred. For all tip widths, the nematic phase is preferred at low densities. }
	\end{figure}

In Fig.\ref{fig:PDSm} we plot the ``phase diagram'' implied by the inequality (\ref{eq:SmecIneq}). The axes are the length fraction $\varphi$ (horizontal) and the reduced tip width $x$ (vertical). The blue shaded region is where (\ref{eq:SmecIneq}) is \textit{not} satisfied, i.e. the nematic is favoured. The unshaded portion is where the smectic becomes preferred over the nematic. From this diagram we can see that, as expected, at low particle densities ($\varphi \ll 1$) the equilibrium state is the nematic. As $\varphi$ is increased, however, this can become unstable to the smectic phase, depending on the particles' tip width. Importantly, only those particles with \textit{sufficiently wide} tips see the smectic phase and those with wider tips make smectic at \textit{lower} $\varphi$. 

\begin{figure}\includegraphics[width=8cm]{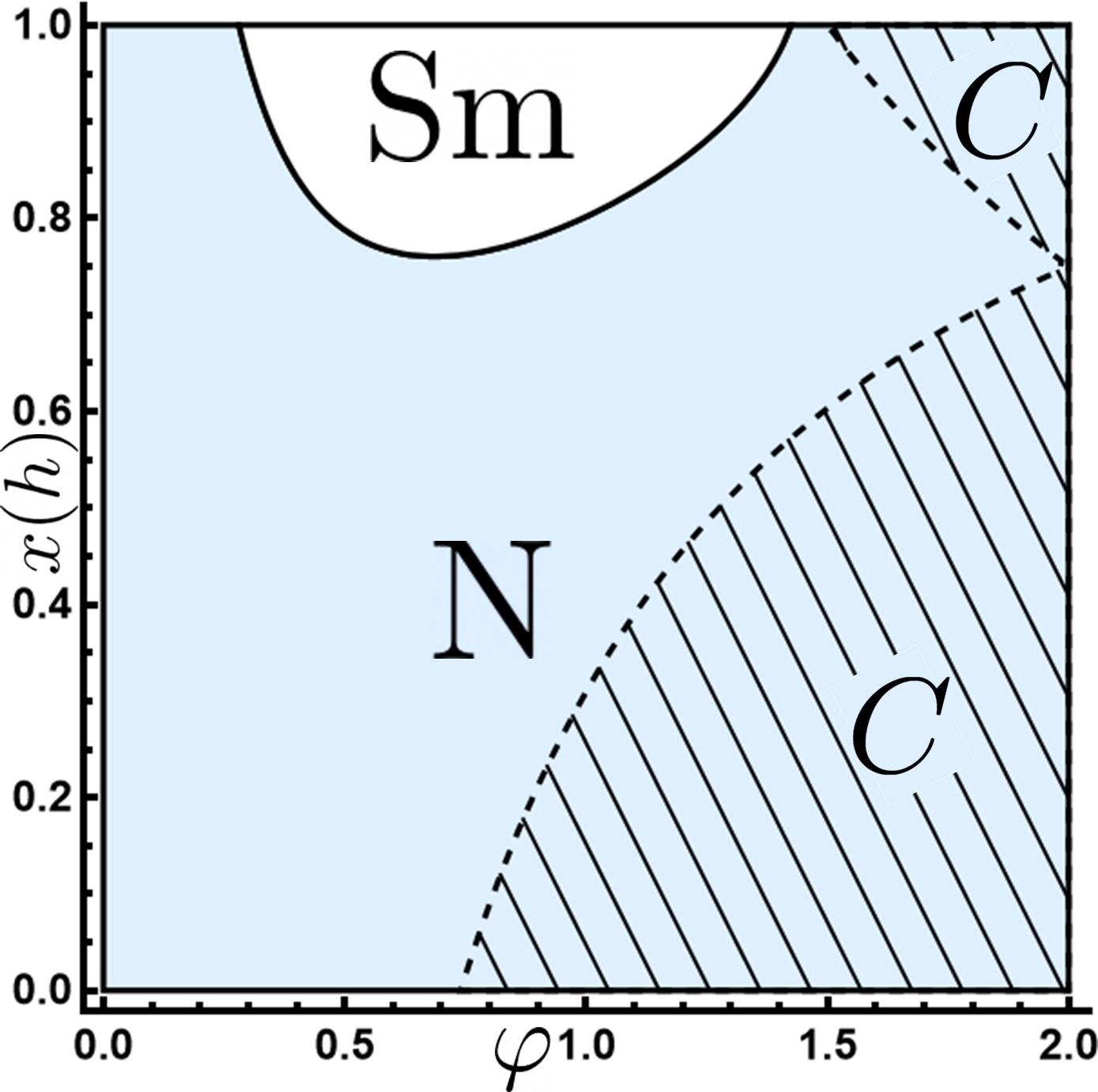}
		\caption{\label{fig:PDCy} The phase diagram as a function of length fraction, $\varphi = N w_0 /L$, and reduced tip width $x(h)= w(h)/w_0$. The region where the nematic is preferred is shaded in blue and the unshaded region is where the smectic is preferred. The hashed regions at high $\varphi$ are the regions where the crystal is preferred over the smectic. This shows that the order of preferred phases as $\varphi$ is increased for particles with thin tips (small $x$) is; nematic - crystal. For those with wider tips it is expected to be; nematic - smectic - crystal. However, while the crystal is preferred over the smectic at high densities, the one dimensional nature of the crystalline order in the model forces it to melt to a nematic.}
	\end{figure}

The upper right hand corner of the phase diagram should also be discussed. This represents the very high density region for wide-tipped particles. Here, the diagram indicates the nematic phase is favoured. This means that, if a smectic phase can be preferred, it only remains until a relatively high density, where the nematic is preferred again. To understand this, let us consider a ``crystal'' phase. This is one where all layers are equally occupied and between each particle on the central layer there are two on the outer layers, one on the upper and one on the lower. We call this arrangement ``crystalline'' as it visually mimics the positional order expected for the crystal. The free energy of such a phase is readily calculated from (\ref{eq:PFFinal})
\begin{equation}
	\beta F_{\text{C}} = - N \left[\log (1 - 2 \varphi x /3) + \frac{1}{3} \log 2 + 1 - \log \rho \right]. 
\end{equation}
This can be used to determine when the crystal is preferred to the smectic, which happens when
\begin{equation}
	\frac{2 - \varphi}{3- 2 \varphi x} < \frac{2}{3} 2^{-3/2}. 
\end{equation}
This condition is added to the phase diagram in Fig.\ref{fig:PDCy}, with the region where the crystal is preferred to the smectic shown hashed. This shows that in the upper right corner of the phase diagram, the smectic phase is beaten by the crystal. Particles with thinner tips also see their, non-existent, smectic phases lose out to the crystal at high densities. This is reasonable because we expect to see the crystal as the phase preferred at highest densities, not the smectic. What is also understandable is that the crystal phase is \textit{never} preferred over the nematic. This is because the crystalline order we have considered is essentially one dimensional, and it is known that fluctuations will always destroy order in one dimension \cite{VanHove1950}. 

In the same spirit as that final point, it is worth reiterating what we have done here. We calculated the free energy in the pure smectic phase and the nematic phase and compared them to each other, finding the regions where one is preferred over the other. We did not strictly show that a transition to a smectic phase occurs in the presence of fluctuations. Inevitably one would find no pure smectic phase in this system because it is, in a sense, one dimensional. Nevertheless, we argue that we have still gained useful insight from our analysis. In particular, demonstrating the role of particle tip geometry in the competition between the nematic and smectic phases. 

\section{Discussion and Conclusions}
\label{sec:Conc}
In this paper, we considered a reduced model for a two-dimensional smectic. The particles were restricted to three equally spaced layers. The outer two represent smectic layers while the central one played the role of a layer interstice. This is essentially an embellishment of the exactly soluble Tonks gas \cite{Tonks1936TheSpheres} and we were able to find the partition function in the isobaric (constant pressure) ensemble in closed form. In the thermodynamic limit, the canonical partition function may be expressed concisely. This allows us to compare the free energies of the analogue to the smectic phase, where the central layer is depleted, and the nematic phase, where all three layers are equally occupied. In the context of this exactly soluble model, we demonstrated that the balance between the entropic gain from not introducing particles between the smectic layers and the subsequent reduction in the number of configurations is what drives the competition between the nematic and smectic phases. We have also understood the role that the geometry of the particles play in tipping this balance one way or the other; wider tips make occupying layer interstices more difficult, and nematics less stable than smectics. These conclusions are consistent with the previous mean-field theory which addressed the affect of tip geometry on smectic formation \cite{King2023} and places those intuitive arguments on firmer footing. It also goes some way to explaining the observation that thinner tipped ellipsoids miss the smectic phase presented by wider tipped sphero-cylinders \cite{Frenkel1991,Frenkel1984,Stroobants1986}.

Of course, the model that we have considered here is extremely simple. Real smectics exist in three dimensions and contain more than three layers. Nevertheless, the simplicity of the model can allow more complicated features to be introduced in a venue where theoretical analysis is still feasible. For instance, interactions \textit{between} the layers may be introduced. This could help refine the treatment presented in the mean-field model for the interaction betwee the $N$-CB alkyl tails, and is also relevant for smectics made from DNA, where the interactions are mediated by ``sticky'' single-stranded DNA sequences dangling between the layers \cite{Gyawali2021}. Interactions other than excluded volume between particles may be introduced the Tonks gas\cite{Koppel1963, Kofke1993}, and this is where the transfer operator method would find its home in our model. 

Mixtures of different particles may also be considered. The most interesting case arises when particles whose shape prevents a smectic are mixed with those promoting one. This serves up a veritable sm\"{o}g\r{a}sbord of questions; at what ratio of smectic preventers to promoters is the smectic lost? Does de-mixing occur and, if so, when? What are the arrangements of particles at intermediate proportions? The interaction between two different particle shapes can be handled relatively straightforwardly by changing the function $w(h)$ and hence the functions $W_k(n)$. Perhaps this simple model system is a good place to start.

 \acknowledgements
I thank Professor Randall D. Kamien for his time spent discussing this work and critical reading the manuscript as well as his continued encouragement. This work was supported by Simons Investigator Grant No. 291825 from the Simons Foundation.
\appendix
\section{Demonstrating the Trick}
\label{app:Trick}
Suppose we are to evaluate the following for large $N$,
\begin{equation}
	T(N) = \prod_{k=0}^{N} \sum_{\substack{n_k \geq 0 \\ \sum_{k=0}^{N} n_k = N}} C(n_k),
\end{equation}
where $C(n_k)$ are some real coefficients. We can remove the constraint on the summation by introducing the delta function $\delta(\sum_{k=0}^{N} n_k - N)$, so that
\begin{equation}
	T(N) = \prod_{k=0}^{N} \sum_{n_k \geq 0 } \delta\left(\sum_{k=0}^{N} n_k - N\right) C(n_k). 
\end{equation}
The delta function is then written as a Fourier integral
\begin{equation}
	\delta\left(\sum_{k=0}^{N} n_k - N\right) = \int_{-\pi}^{\pi} \frac{d\theta}{2\pi} e^{- i N \theta} \exp\left(i \sum_{k=0}^{N} n_k \theta \right),
\end{equation}
where we have used that $N$ and all the $n_k$ are integers. This separates out the sums on $n_k$, at the cost of introducing the integral over $\theta$
\begin{equation}
	T(N) = \int_{-\pi}^{\pi} \frac{d\theta}{2\pi} e^{- i N \theta} \prod_{k=0}^{N} \sum_{n_k \geq 0 }^{N} C(n_k) e^{i \theta n_k}.
\end{equation}
Now let us suppose that the coefficients $C(n_k)$ decay sufficiently quickly to allow the upper limit of the sum on $n_k$ to be taken to $\infty$, which will have negligible error for $N \to \infty$, and for the following function to exist
\begin{equation}
	f(z) = \sum_{n=0}^{\infty} C(n) z^n.
\end{equation}
This gives, upon exponentiating the product,
\begin{equation}
	T(N) = \int_{-\pi}^{\pi} \frac{d\theta}{2\pi} \exp\left(N \log f(e^{i\theta}) - i N \theta\right).
\end{equation}
If all is well, the function $f(z)$ will allow this integral to be taken for $N \gg 1$ by standard methods (e.g. saddle point or steepest decent). 

As a concrete example, let us use this method for the multinomial theorem. This requires computing 
\begin{equation}
	T_M(N) = N! \prod_{k=0}^{N} \sum_{\substack{n_k \geq 0 \\ \sum_{k=0}^{N} n_k = N}} \frac{1}{n_k!}.
\end{equation}
Imposing the delta function and writing it in fourier transform, just as before, yields
\begin{equation}
	T_M(N) = N! \int_{-\pi}^{\pi} \frac{d\theta}{2\pi} e^{- i N \theta} \prod_{k=0}^{N} \sum_{n_k =0}^{N} \frac{e^{i \theta n_k}}{n_k!}.
\end{equation}
In this case we can take $N \to \infty$ on the summation to obtain
\begin{equation}
	\label{eq:AppInt}
	T_M(N) = N! \int_{-\pi}^{\pi} \frac{d\theta}{2\pi} e^{- i N \theta} \prod_{k=0}^{N} \exp\left(e^{i \theta}\right) = N! \int_{-\pi}^{\pi} \frac{d\theta}{2\pi} \exp\left(N e^{i \theta} - i N \theta \right).
\end{equation}
The asymptotic behaviour for large $N$ of the $\theta$ integral can be studied in detail in lots of ways. For our purposes, it is sufficient to notice that the maximum of the real part of the function in the exponent occurs when $\theta=0$ and so, to leading order, the integral is given by the value of the integrand there
\begin{equation}
	T_M(N) \sim N! e^N \sim N^N.
\end{equation}
This is the result for $T_M(N)$ required by the multinomial theorem which demonstrates how this trick can be used to obtain consistent results for $N \gg 1$. 

For completeness we should note that, upon changing variables to $z = e^{i \theta}$, applying the residue theorem to the integral in (\ref{eq:AppInt}) shows that $T_M(N) = N^N$ \textit{precisely}. 
\section{Manipulating \texorpdfstring{$\Phi$}{TEXT}}
\label{app:ExpGenL}
In this appendix we show how $\Phi(x,y)$, as given in (\ref{eq:PhiDefInf}), can be written as the infinite sum of modified Bessel functions (\ref{eq:PhiBessel}). Let us first define
\begin{equation}
	S_n(z)\equiv\sum_{m=0}^{n} \frac{(n+m)!}{(m!)^2} z^m,
\end{equation}
so that $\Phi$ can be written
\begin{equation}
	\Phi(y,z) = \sum_{n=0}^{\infty} \frac{1}{(n!)^2} S_n(z) y^n.
\end{equation}
It is useful to split $S_n$ into two parts
\begin{equation}
	S_n(z) = \sum_{m=0}^{\infty} \frac{(n+m)!}{(m!)^2} z^m - \sum_{m=n+1}^{\infty} \frac{(n+m)!}{(m!)^2} z^m \equiv S_{\infty}(z) - R_n(z).
\end{equation}
We shall consider $S_{\infty}(z)$ and $R_n(z)$ in turn. 
\subsection{\texorpdfstring{$S_{\infty}(z)$}{TEXT}}
We aim to find the differential equation satisfied by $S_{\infty}(z)$. Its first two derivatives are
\begin{subequations}
	\begin{equation}
		S_{\infty}'(z) = \sum_{m=0}^{\infty} \frac{(n+m)!}{(m!)^2} m z^{m-1},
	\end{equation}
and
	\begin{equation}
	S_{\infty}''(z) = \sum_{m=0}^{\infty} \frac{(n+m)!}{(m!)^2} m (m-1) z^{m-2}.
\end{equation}
\end{subequations}
Multiplying the second derivative by $z$ and adding it to the first gives
\begin{equation}
	\sum_{m=0}^{\infty} \frac{(n+m)!}{(m!)^2} m^2 z^{m-1} = \sum_{m=0}^{\infty} \frac{(n+m+1)!}{[(m+1)!]^2} (m+1)^2 z^{m} = \sum_{m=0}^{\infty} \frac{(n+m)!}{(m!)^2} (n+m+1)z^{m}.
\end{equation}
This is almost $S_{\infty}(z)$ itself, save for the errant $m$ in the numerator of the final summand. This may be removed by subtracting $z S_{\infty}'(z)$. Hence $S_{\infty}(z)$ satisfies
\begin{equation}
	z S_{\infty}''(z) + (1-z)S_{\infty}'(z) - (n+1)S_{\infty}(z) = 0.
\end{equation}
The two solutions to this equation are a Laguerre polynomial of negative index $L_{-(n+1)}(z)$  and the confluent hypergeometric function $\mathcal{U}(n+1,1,z)$ \footnote{see Chap. 13 of \cite{Abramowitz1964HandbookTables}}. The Laguerre polynomial is finite at $z=0$, in fact $L_{-(n+1)}(0) = 1$, whereas $\mathcal{U}$ diverges. From its definition, we have $S_{\infty}(0) = n!$. Therefore
\begin{equation}
	S_{\infty}(z) = n! L_{-(n+1)}(z) = n! e^{z} L_n (-z),
\end{equation}
where we have made use of the identity $L_{-l}(z) = e^z L_{l-1}(-z)$ to make the index of $L$ positive. The contribution of this term to $\Phi$ can then be written as
\begin{equation}
	\label{eq:PhiInf}
	\Phi_{\infty}(y,z) \equiv \sum_{n=0}^{\infty} \frac{1}{(n!)^2} S_{\infty}(z) y^n = e^{z}\sum_{n=0}^{\infty} \frac{y^n}{n!}L_n(-z).
\end{equation}
The summation is then recognised as the exponential generating function for the Laguerre polynomials. This may be found in closed form directly from their ordinary generating using the Borel transform \cite{Bender1978}. Naming the required generating function
\begin{equation}
	G_B(y,z) = \sum_{n=0}^{\infty} \frac{y^n}{n!}L_n(z).
\end{equation}
The ordinary generating function has a well known closed form \cite{Lebedev1972}
\begin{equation}
	\label{eq:GenLCF}
	G(y,z)= \sum_{n=0}^{\infty} y^n L_n(z) = \frac{1}{1-y} \exp\left(-\frac{yz}{1-y}\right).
\end{equation}
We shall relate this to $G_B$ using the Borel transform of a series. First we introduce a factor of unity to the above using the integral representation of the Gamma function
\begin{equation}
	G(y,z) = \sum_{n=0}^{\infty} y^n L_n(z) \frac{1}{n!} \int_{0}^{\infty} dx \ x^n e^{-x}. 
\end{equation} 
Or, upon commuting the summation and integral,
\begin{equation}
	G(y,z) = \int_{0}^{\infty} dx \ e^{-x} \sum_{n=0}^{\infty} \frac{(x y)^n}{n!}L_n(z),
\end{equation}
where we identify the summation as $G_B$
\begin{equation}
	G(y,z) = \int_{0}^{\infty} dx \ e^{-x} G_B(xy,z).
\end{equation}
Next, we introduce the variable $t = x y$
\begin{equation}
	G(y,z) = \frac{1}{y}\int_{0}^{\infty} dt \ e^{-t/y} G_B(t,z).
\end{equation}
Working in terms of $s = 1/y$ reveals that $G$ is related to $G_B$ by the Laplace transform, 
\begin{equation}
	\int_{0}^{\infty} dt \ e^{-s t} G_B(t,z) = \frac{1}{s} G(1/s , z) = \frac{1}{s-1} \exp\left(- \frac{z}{y-1}\right), 
\end{equation}
where we have used (\ref{eq:GenLCF}). To obtain $G_B$ we need to invert the Laplace transform. This may be done using standard tables \footnote{see Chap. 29 of \cite{Abramowitz1964HandbookTables}}, with the result 
\begin{equation}
	G_B(y,z) = e^{y} J_0 (2 \sqrt{y z}),
\end{equation}
where $J_0(x)$ is the zeroth order Bessel function of the first kind. Finally we have 
 \begin{equation}
 		\label{eq:PhiInfDone}
 	\Phi_{\infty}(y,z) = e^z G_B(y,-z) =  e^{y + z} I_0(2 \sqrt{y z}),
 \end{equation}
where we have needed the relationship $J_0(i x) = I_0(x)$ between $J_0$ and its modified sibling.
\subsection{\texorpdfstring{$R_{n}(z)$}{TEXT}}
We now need to handle $R_n(z)$ and its contribution 
\begin{equation}
	\Phi_{R}(y,z) =  \sum_{n=0}^{\infty} \frac{1}{(n!)^2} R_n(z) y^n.
\end{equation}
To obtain a convenient form of $R_n(z)$, we have to resort to generalised hypergeometric functions. We can write 
\begin{equation}
	R_n(z) = z^{n+1} \sum_{k=0}^{\infty} \frac{(2n+ 1 + k)!}{[(n+1+k)!]^2} z^k,
\end{equation}
and hence, by consulting tables, e.g. section \href{https://dlmf.nist.gov/16.2}{16.2} of the Digital Library of Mathematical Functions (DLMF) \cite{NIST:DLMF}, we have
\begin{equation}
	R_n(z) = \frac{(2n+1)!}{[(n+1)!]^2} z^{n+1} {}_2 F_2 (1 , 2n+2 ; n+2 , n+2 ; z).
\end{equation}
We can relate the generalised hypergeometric function ${}_2 F_2$ to its lower order relative ${}_1 F_1 $ by the integral identity \cite[\href{https://dlmf.nist.gov/16.5}{16.5}]{NIST:DLMF}
\begin{equation}
 {}_2 F_2 (1 , 2n+2 ; n+2 , n+2 ; z) =	\frac{(n+1)!}{n!} \int_{0}^{1} dt \ (1-t)^n {}_1 F_1 (2n+2 ; n+2 ; z t).
\end{equation} 
The function ${}_1 F_1 (2n+2 ; n+2 ; z t)$ admits the following integral representation of Mellin-Barnes type \cite[\href{https://dlmf.nist.gov/13.4}{13.4}]{NIST:DLMF}
\begin{equation}
	{}_1 F_1 (2n+2 ; n+2 ; z t) = \frac{(n+1)!}{(2n+1)!}\int_{\gamma - i \infty}^{\gamma + i \infty} \frac{ds}{2\pi i} \frac{\Gamma(-s)\Gamma(s + 2n +2)}{\Gamma(s+n+2)} (- z t)^{s}
\end{equation}
where the constant $\gamma$ is chosen so that the contour runs to the left of the poles of $\Gamma(-s)$ on to the right of those of $\Gamma(s + 2n +2)$. This lets us write $R_n(z)$ as two integrals
\begin{equation}
	R_n(z) = \frac{z^{n+1}}{n!} \int_{0}^{t} dt  \  \int_{\gamma - i \infty}^{\gamma + i \infty} \frac{ds}{2\pi i} \frac{\Gamma(-s)\Gamma(s + 2n +2)}{\Gamma(s+n+2)} (1-t)^n (- z t)^{s}.
\end{equation}
The integral over $t$ is straightforward
\begin{equation}
	\int_{0}^{1} dt\ (1-t)^n (-z t)^s = \frac{\Gamma(s+1)}{\Gamma(s + n + 2)} n! (-z)^s.
\end{equation}
 Hence we have, 
\begin{equation}
	R_n(z) = z^{n+1} \int_{\gamma - i \infty}^{\gamma + i \infty} \frac{ds}{2\pi i} \frac{\Gamma(-s)\Gamma(s+1)\Gamma(s + 2n +2)}{[\Gamma(s+n+2)]^2} (-z)^s .
\end{equation}
Thus, to find $\Phi_R$, we require
\begin{equation}
	\Phi_{R}(y,z) =  z \sum_{n=0}^{\infty}\int_{\gamma - i \infty}^{\gamma + i \infty} \frac{ds}{2\pi i} \frac{\Gamma(-s)\Gamma(s+1)\Gamma(s + 2n +2)}{[\Gamma(s+n+2)]^2 (n!)^2} (-z)^s (y z)^n.
\end{equation}
Pulling the summation past the integral gives, 
\begin{equation}
	\Phi_{R}(y,z) =  z \int_{\gamma - i \infty}^{\gamma + i \infty} \frac{ds}{2\pi i} \Gamma(-s)\Gamma(s+1) (-z)^s \sum_{n=0}^{\infty} \frac{\Gamma(s + 2n +2)}{[\Gamma(s+n+2)]^2 (n!)^2}(y z)^n.
\end{equation}
Careful comparison to general definitions \cite[\href{https://dlmf.nist.gov/16.2}{16.2}]{NIST:DLMF} reveals the sum to be a yet more complicated hypergeometric function,
\begin{equation}
	\label{eq:app2F3}
	\sum_{n=0}^{\infty} \frac{\Gamma(s + 2n +2)}{[\Gamma(s+n+2)]^2 (n!)^2}(y z)^n = \frac{1}{\Gamma(s+2)} {}_2 F_3(1 + s/2 , 3/2 + s/2 ; 1 , 2 + s, 2 + s; 4 y z). 
\end{equation}
Fortunately, this may be reduced to the product of two simpler functions by the identity \cite[\href{https://dlmf.nist.gov/16.12}{16.12}]{NIST:DLMF}
\begin{equation}
	{}_2 F_3((a+b)/2 , (a+b -1)/2 ; a ,b, a+b-1; 4 x) = {}_0 F_1( ; 1 ; x) {}_0 F_1( ; b ; x).
\end{equation}
From (\ref{eq:app2F3}) we identify, $a=1$, $b= s+2$ and $x = y z$. This relation is supremely advantageous because the functions ${}_0 F_1$ are directly related to modified Bessel functions by \cite[\href{https://dlmf.nist.gov/13.6}{13.6}]{NIST:DLMF}
\begin{equation}
	{}_0 F_1( ; a+1 ; x) = \frac{\Gamma(a+1)}{(x/2)^a}I_{a}(x^2/4).
\end{equation}
Hence we have reduced the sum to a product of two Bessel functions
\begin{equation}
	\sum_{n=0}^{\infty} \frac{\Gamma(s + 2n +2)}{[\Gamma(s+n+2)]^2 (n!)^2}(y z)^n = (\sqrt{y z})^{-(s+1)} I_{0}(2 \sqrt{yz})I_{s+1}(2 \sqrt{yz}).
\end{equation}
We now have the following contour integral with which to contend, 
\begin{equation}
	\Phi_{R}(y,z) =  \sqrt{\frac{z}{y}} I_0(2\sqrt{y z}) \int_{\gamma - i \infty}^{\gamma + i \infty} \frac{ds}{2\pi i} \ \left(-\sqrt{\frac{z}{y}}\right)^{s} \Gamma(-s)\Gamma(s+1) I_{s+1}(2\sqrt{y z}).
\end{equation}
To employ the residue theorem, we need to close the contour. The contour should be closed in such a way as to pick up the residues from the poles of $\Gamma(-s)$ at $s= 0 , 1 , 2 $ etc. To do this, we need to add the integral around the infinite semi-circle which sweeps from the negative to positive imaginary axes. All being well, this additional integral is zero so $\Phi_R$ is just the sum of the residues. This is indeed the case. To see this we set $s$ to be very large, positive, and real. The magnitude of the integrand may then be deduced from the asymptotics of the Bessel function for large orders \cite{Abramowitz1964HandbookTables}, since the two Gamma functions cancel each other, 
\begin{equation}
	I_{s+1}(x) \sim \frac{1}{\sqrt{2 \pi s}}\frac{e^s x^s}{2 s^s}.
\end{equation}
The factor of $s^s$ in the denominator means that the integrand, approaches zero for large $s$. Thus, we may close the contour, and the residue theorem gives us
\begin{equation}
	\Phi_{R}(y,z) =  \sqrt{\frac{z}{y}} I_0(2\sqrt{y z}) \sum_{n=0}^{\infty} \text{Res}\left[\left(-\sqrt{\frac{z}{y}}\right)^{s} \Gamma(-s)\Gamma(s+1) I_{s+1}(2\sqrt{y z}) ; s = n\right].
\end{equation}
The residue of $\Gamma(-s)$ at integer $n$ is $(-1)^n (n!)^{-1}$, and so we have
\begin{equation}
	\Phi_{R}(y,z) =   I_0(2\sqrt{y z}) \sum_{n=0}^{\infty} \left(\sqrt{\frac{z}{y}}\right)^{n+1} I_{n+1}(2\sqrt{y z})
\end{equation}
Shifting the lower limit of the sum to $n=1$, and adding this to $\Phi_{\infty}$, we have finally obtained (\ref{eq:PhiBessel}) in the main text. 
\begin{equation}
	\Phi(y,z)  = I_0(2 \sqrt{y z}) \left[e^{y+z} - \sum_{n=1}^{\infty} \left(z/y\right)^{n/2} I_n (2 \sqrt{y z}) \right],
\end{equation}
It is worth noting that that the infinite summation in the brackets appears closely related to the generating function for the modified Bessel functions, for which there is a closed form \cite{Abramowitz1964HandbookTables}. That is written 
\begin{equation}
	 \label{eq:BessGen}
	\sum_{n=-\infty}^{\infty} x^n I_{n}(y) = e^{y(x + x^{-1})/2}. 
\end{equation}
Our summation is the ``one-sided'' equivalent. Despite this seduction, it appears that there is no especially useful way to use the closed form (\ref{eq:BessGen}) to simplify (\ref{eq:PhiBessel}) further.
\subsection{A curious identity}
A pleasing consequence of this labour is a neat expression for the value $\Phi(1,1)$. We have shown that this should be
\begin{equation}
	\Phi(1,1) = I_0(2)\left(e^2 - \sum_{n=1}^{\infty} I_n(2)\right).
\end{equation}
Here we can employ the identity 
\begin{equation}
	e^{z} = I_0(z) + 2\sum_{n=1}^{\infty} I_{n}(z),
\end{equation}
to find the appealing relation
\begin{equation}
	\sum_{n=0}^{\infty}\sum_{m=0}^{n} \frac{(n+m)!}{(n!)^2 (m!)^2} = \frac{1}{2}I_0(2) \left(e^2 - I_0(2)\right).
\end{equation}
\section{Maximising \texorpdfstring{$\Phi$}{TEXT}}
\label{app:PhiMax}
Here we show that the maximum of $\Phi(e^{i \theta - \alpha},e^{i \eta})$ occurs when $\theta = \eta = 0$. From the definition of $\Psi$ in (\ref{eq:PF3}), this is enough to show that $|\Psi|$ is also maximal there. 

Let us consider fixing $y$ at a given real value, say $y = c^2/4$ for convenience, and varying $z$ around the unit semi-circle in the complex plane; i.e. $z = e^{i \theta}$ with $\theta \in [-\pi , \pi]$. We would like to find which value of $\theta$ maximises the absolute value of 
 \begin{equation}
 	\label{eq:PhiBessRef}
 	\Phi(c^2/4,e^{i \theta})  = I_0(c e^{i \theta/2}) \left[e^{c^2/4+e^{i \theta}} - \sum_{n=1}^{\infty} (2/c)^n e^{i n \theta/2} I_n (2c e^{i \theta/2}) \right].
 \end{equation}
First we note that the magnitude of the $e^{i n \theta/2}$ factor is unity. Then we write each Bessel function as 
\begin{equation}
	I_n(c e^{i \theta/2}) = \int_{0}^{\pi} \frac{dq}{\pi} \cos (n q)  \exp\left[c \cos q \cos \theta/2+ i c \cos q \sin \theta/2\right].
\end{equation}
From which it is clear that the magnitude of the integral is supressed away from $\theta = 0$, where it is maximal. Next we consider the first term in the brackets of (\ref{eq:PhiBessRef}). The relevant $\theta$ dependent part can be written
\begin{equation}
	e^{e^{i\theta}} = \exp\left[\cos(\theta) + i \sin \theta)\right]. 
\end{equation} 
Again it is easy to see that this reaches its maximum absolute value at $\theta = 0$. We may therefore conclude 
\begin{equation}
	\max_{\theta \in [-\pi, \pi]}|\Phi(c^2/4, e^{i\theta})| = \Phi(c^2/4, 1).
\end{equation}
Applying this reasoning when $z$ is fixed and $y$ is varied allows us to conclude that 
\begin{equation}
	\max_{\theta , \eta \in [-\pi, \pi]}|\Phi(e^{i \theta - \alpha}, e^{i \eta})| = \max_{\theta , \eta \in [-\pi, \pi]}|\Phi(e^{i \eta - \alpha}, e^{i \theta})| = \Phi(e^{-\alpha}, 1).
\end{equation}
This can be confirmed by plotting $\Phi(y, z)$ numerically, as is done in Fig.(\ref{fig:PhiPlot}).
\bibliography{references}

\end{document}